\documentclass[aps,pre,a4paper,twocolumn]{revtex4-2}

\usepackage{psfrag}
\usepackage{amssymb,amsmath,amsthm}
\usepackage[dvips]{graphicx}
\usepackage{verbatim}
\usepackage{color}
\usepackage{siunitx}

\begin{document}

\title{Mediated interactions between rigid inclusions in two-dimensional elastic or fluid films}

\date{\today}

\author{S. K. Richter }

\email{sonja.richter@ovgu.de} 

\author{A. M. Menzel}

\email{a.menzel@ovgu.de} 
    
\affiliation{Institut f\"ur Physik, Otto-von-Guericke-Universit\"at Magdeburg, Universit\"atsplatz 2, 39106 Magdeburg, Germany}

\begin{abstract}
Interactions between rigid inclusions in continuous three-dimensional linearly elastic solids and low-Reynolds-number viscous fluids have largely been quantified during the past. Prime example systems are given by functionalized elastic composite materials or fluid colloidal suspensions. Here, we address the significantly less frequently studied situation of rigid inclusions in two-dimensional elastic or low-Reynolds-number fluid films. We concentrate on the situation in which disk-like inclusions remain well separated from each other and do not get into contact. Specifically, we demonstrate and explain that the logarithmic divergence of the associated Green's function is removed in the absence of net external forces on the inclusions, in line with physical intuition. For instance, this situation applies when only pairwise mutual interactions between the inclusions prevail. Our results will support, for example, investigations on membranes functionalized by appropriate inclusions, both of technical or biological origin, or the dynamics of active microswimmers in appropriately prepared thin films. 
\end{abstract}

\maketitle


\section{Introduction} \label{sec:intro}

Already from a practical point of view, functionalization of solids and fluids by more or less rigid inclusions is of paramount interest. We think, for example, of elastic composite materials that through externally addressable rigid inclusions may serve as soft actuators \cite{hines2017soft, fischer2020towards, bose2021magnetorheological} or of colloidal suspensions \cite{batchelor1972determination, dhont1996introduction, poon2004colloids} that represent materials of everyday usage such as paint. A more specific area of recent interest extends to the world of active microswimmers self-propelling through a liquid \cite{elgeti2015physics, zottl2016emergent} and thus forming a type of active suspension, for which characteristic properties such as reduced viscosities may be expected \cite{sokolov2009reduction, saintillan2010dilute, gachelin2013non}. These basic examples motivate a theoretical characterization of such set-ups, particularly concerning the mutual interactions between the inclusions that may be mediated by the surrounding environment. 

More in detail, if a force or torque is exerted on one inclusion, it will be transmitted to some extent to the surrounding medium. The medium is set into motion and gets displaced. Since the other inclusions are surrounded by the medium as well, their configuration is likewise affected. Overall, the total configuration of the inclusions is substantially coupled by the enclosing environment. 

A lot of effort has been spent over the past decades to calculate and quantify such mutual interactions. This concerns both elastic solids \cite{phan1994load, kim1995faxen, schopphoven2019elastic, puljiz2017pre, puljiz2019pre-higher-order} and viscous fluids \cite{mazur1982many, kim2013microhydrodynamics, dhont1996introduction, hoell2019multi} as surrounding media, where in the latter context a focus is on incompressible liquids subject to low-Reynolds-number flows. To be able to perform analytical calculations, a major focus was on bulk states far away from any boundaries and rigid spherical inclusions experiencing no-slip conditions for the surrounding medium on their surfaces. Deviations from these limitations are possible to some extent and have been addressed as well \cite{blake1971note, liron1976stokes, liron1978stokes, squires2000like, zottl2012nonlinear, menzel2017force, daddi2017hydrodynamic, daddi2018state}. Moreover, the majority of analytical approaches considers identical inclusions of definite surface-to-surface distance from each other. Then an iterative analytical procedure becomes possible that corresponds to an expansion inverse in the center-to-center inclusion distance. The regime of quantitative validity of such an expansion depends on the situation and tolerance in deviations, but minimal surface-to-surface distances of one inclusion radius or diameter is typically reasonable. Below, we will rely on corresponding simplifying assumptions as well. 

For obvious reasons, most of the systems addressed so far were genuinely three-dimensional. However, there are situations in which two-dimensional approaches become appropriate. For example, the equations of linear elasticity of thin elastic membranes can effectively be reduced to two dimensions \cite{landau1986theory}. Similarly, the equations describing the dynamics in thin fluid films can be reduced to the two-dimensional inplane spatial variation of the film thickness \cite{oron1997long}. We note that, nevertheless, a truly two-dimensional approach does not necessarily imply that the system is infinitely thin. Instead, three-dimensional systems that behave completely homogeneously concerning the third dimension, that is, there is no spatial dependence on the third dimension, can frequently be treated in an effectively two-dimensional way. Then, at least for illustration of such three-dimensional systems, we may think of infinitely extended parallel cylinders as rigid inclusions in a surrounding elastic or fluid medium. 

Our scope in this work is therefore to provide the framework of mediated interactions between rigid inclusions in a surrounding soft elastic solid or an incompressible low-Reynolds-number fluid in a two-dimensional setting. Within the two-dimensional framework, we refer to the inclusions as disk-like. We assume these disks to be subject to additional forces and/or torques that are not directly exerted on them through their surfaces by the surrounding medium. Instead, they result, for instance, from pairwise magnetic interactions between the inclusions. This drives the resulting overall configurational changes. 

Specifically in two dimensions, a logarithmic divergence of the associated Green's function describing the response of the medium to an internal force center emerges \cite{phan-thien-paper-elastic-2D}. This logarithmic divergence has been the subject to previous discussions \cite{Squires2006}. Physically, it implies that a resting two-dimensional system, if not explicitly held fixed at its boundaries, cannot sustain in the linear regime a net force, even if the force is applied at only a single point and even if the system is infinitely extended. The authors of Ref.~\onlinecite{proudman_pearson_1957} argued that in such a case the framework of the linear theory breaks down. Instead, in their fluid environment, the nonlinear convective term of the Navier--Stokes equations becomes important. Here, we demonstrate explicitly that the logarithmic divergence only emerges when a net external force is imposed on the inclusions. If all forces on the inclusions sum up to zero, which is the case, for instance, for conservative pairwise interactions, then there do not appear any divergences. Nevertheless, in general, the mutual mediated interactions between the inclusions are more long-ranged in two dimensions than in three-dimensional systems. 

We proceed as follows.  First, the basic underlying equations are briefly repeated in Sec.~\ref{sec:basic}. The corresponding Green's function containing a logarithmic divergence is listed in Sec.~\ref{sec:Green}. Afterwards, in Sec.~\ref{sec:force} and in Sec.~\ref{sec:torque}, the translational and rotational response of the surrounding medium to a force and torque on an individual disk-like inclusion are summarized, respectively. Then, corresponding Fax\'{e}n relations, that is, the translational, rotational, and stresslet types of response of a disk-like inclusion to a displacement or flow of the surrounding medium are derived in Sec.~\ref{sec:Faxen}. Next, in Secs.~\ref{sec:matrix} and \ref{sec:interaction_two}, we calculate the translation--translation, rotation--translation, translation--rotation, and then rotation--rotation couplings via the surrounding medium between individual inclusions and cast them into matrix form. In Sec.~\ref{sec:interaction_three}, we consider three-body interactions. Finally, we demonstrate in Sec.~\ref{sec:ln-problem} how the logarithmic divergence arising, for example, in Sec~\ref{sec:Green} is naturally removed, and we compare our results to the ones of a three-dimensional setup in Sec.~\ref{sec:2Dvs3D}. We conclude in Sec.~\ref{sec:conclusion}.

\section{Basic equations for the medium}\label{sec:basic}

Our basis is given by the Navier--Cauchy equations of linear elasticity \cite{cauchy1828exercices} for an isotropic, homogeneous, infinitely extended, continuous medium
\begin{equation}\label{eq:NCE}
	\nabla^2 \mathbf{u}\left(\mathbf{r}\right) + \frac{1}{1-2\nu} \nabla\nabla \cdot \mathbf{u}\left(\mathbf{r}\right) = -\frac{1}{\mu}\mathbf{f}_{\mathrm{b}}(\mathbf{r}),
\end{equation}
here interpreted in two spatial dimensions. In these equations, $\mathbf{u}(\mathbf{r})$ denotes the displacement field that quantifies the distance by which the individual volume elements at positions $\mathbf{r}$ of the elastic material are displaced. $\mu$ sets the elastic (shear) modulus and $-1<\nu<1/2$ the Poisson ratio connected to the compressibility of the material. Incompressible systems are identified by $\nu=1/2$, $\mathbf{f}_{\mathrm{b}}(\mathbf{r})$ describes the field of bulk force density. 

Additionally, embedded within this elastic medium, we consider $N$ rigid disk-like inclusions of radius $a$ at center positions $\mathbf{r}_i$ ($i=1,...,N$). No-slip boundary conditions apply on their circumferences. That is, if an inclusion is displaced as given by a vector $\mathbf{U}_i$ and/or rotated by a rotation vector $\mathbf{\Omega}_i$ ($i=1,...,N$), the elastic material on its surface is displaced accordingly. 

In our presentation below, we mainly refer to these equations of linear elasticity. Yet, we keep in mind that because of the formal analogy to the Stokes equations \cite{dhont1996introduction, kim2013microhydrodynamics}
\begin{equation}\label{eq:Stokes}
	\nabla^2 \mathbf{v}(\mathbf{r})=\frac{1}{\eta}\big(\nabla p\left(\mathbf{r}\right) - \mathbf{f}_b\left(\mathbf{r}\right)\big),
\end{equation} 
the results for viscous incompressible fluid systems under low-Reynolds-number conditions are derived simultaneously. Simply, $\mathbf{u}(\mathbf{r})$ needs to be replaced by the fluid flow field $\mathbf{v}(\mathbf{r})$, analogously the displacements $\mathbf{U}_i$ and rotations $\mathbf{\Omega}_i$ derived below for the inclusions by their velocities $\mathbf{V}_i$ and angular velocities $\mathbf{W}_i$, respectively, and we set $\nu=1/2$. Requiring $\nabla\cdot\mathbf{v}(\mathbf{r},t)=0$ explicitly for incompressible fluids, the term containing the pressure field $p(\mathbf{r})$ in Eq.~(\ref{eq:Stokes}) does not explicitly influence the results.

\section{Green's function}\label{sec:Green}
First, we address Eq.~(\ref{eq:NCE}) for a bulk point force density $\mathbf{f}_{\mathrm{b}}(\mathbf{r})=\mathbf{F}_0\delta(\mathbf{r}-\mathbf{r}_0)$ acting at position $\mathbf{r}_0$, where $\mathbf{F}_0$ sets the strength and direction of the force while $\delta$ denotes the Dirac delta function. Using the Green's function formalism, the solution can then be written as
\begin{equation}\label{eq:uGF}
	\mathbf{u}\left(\mathbf{r}\right) = \underline{\mathbf{G}}\left(\mathbf{r},\mathbf{r}_0\right)\cdot\mathbf{F}_0.
\end{equation}
In two dimensions, the corresponding Green's function was introduced as \cite{phan-thien-paper-elastic-2D}
\begin{equation}\label{eq:Greens}
	\underline{\mathbf{G}}\left(\mathbf{r}\right) = \frac{1}{8\pi\left(1-\nu\right)\mu}\left[-\left(3-4\nu\right)\ln r \hat{\underline{\mathbf{I}}}+\frac{\mathbf{r}\mathbf{r}}{r^2}\right],
\end{equation}
where $\hat{\underline{\mathbf{I}}}$ denotes the unit matrix and $r=|\mathbf{r}|$. This expression can be confirmed by inserting Eqs.~(\ref{eq:uGF}) and (\ref{eq:Greens}) into Eq.~(\ref{eq:NCE}). At first glance, it seems problematic that the first term of the Green's function diverges for $r\rightarrow 0$. We will discuss this aspect later in Sec.~\ref{sec:ln-problem} and show that a consistent description arises from our consideration. 

Consequently, the displacement field $\mathbf{u}(\mathbf{r})$ resulting from the influence of a disk-shaped inclusion reads
\begin{equation}\label{eq:uGf_int}
	\mathbf{u}(\mathbf{r})=\int_{\partial S} \underline{\mathbf{G}}\left(\mathbf{r}-\mathbf{r'}\right)\cdot\mathbf{f}(\mathbf{r'})\, \text dC'\ ,
\end{equation}
where $\partial S$ refers to the circumference of the disk and $\text dC'$ denotes the corresponding line element. $\mathbf{f}(\mathbf{r'})$ quantifies the force per length along this circumference that the disk exerts on the surrounding elastic medium.

We Taylor-expand the $ij$-th component of the Green's function $\underline{\mathbf{G}}(\mathbf{r}-\mathbf{r}')$ in $\mathbf{r}'$ as
\begin{equation}\label{eq:G_exp}
	G_{ij}(\mathbf{r}-\mathbf{r}')=\sum_{n=0}^{\infty}\frac{(-1)^n}{n!}(\mathbf{r}'\cdot\nabla)^n G_{ij}(\mathbf{r}).
\end{equation}
Next, we insert Eq.~(\ref{eq:G_exp}) into Eq.~(\ref{eq:uGf_int}), which results in
\begin{align}\label{eq:u_series}
	u_{i}(\mathbf{r})&=\sum_{n=0}^{\infty}\frac{(-1)^n}{n!}\int_{\partial S}  \text dC'\ (\mathbf{r}'\cdot\nabla)^n G_{ij}(\mathbf{r}) f_{j}(\mathbf{r'}) \nonumber\\
	&= G_{ij}(\mathbf{r})F_j-\frac{\partial  G_{ij}(\mathbf{r})}{\partial r_k}D_{jk}+...\ .
	\end{align}
Focusing on the first two terms in this expansion, we here defined
\begin{equation}
	F_j=\int_{\partial S} \text{d}C'\ f_j(\mathbf{r}')
\end{equation}
and
\begin{equation}
	D_{jk}=\int_{\partial S} \text{d}C'\ f_j(\mathbf{r}')r_k'.
\end{equation} 
Consequently, $\mathbf{F}$ is the total force that the inclusion exerts on its surrounding. Concerning $\underline{\mathbf{D}}$, we split it in into a symmetric part $\underline{\mathbf{S}}$ of components
\begin{equation}\label{eq:stresslet}
	S_{jk}=\frac{1}{2}\int_{\partial S} \text{d}C'\ [f_j(\mathbf{r}')r_k'+f_k(\mathbf{r}')r_j']
\end{equation} and an antisymmetic part $\underline{\mathbf{T}}$ of components
\begin{equation}
	T_{jk}=\frac{1}{2}\int_{\partial S} \text{d}C'\ [f_j(\mathbf{r}')r_k'-f_k(\mathbf{r}')r_j']
\end{equation}
\cite{kim2013microhydrodynamics,dhont1996introduction}. Upon addition they result back in $\underline{\mathbf{D}}$.  Using $\underline{\mathbf{T}}$, we define the components of the torque $\mathbf{T}$ as
\begin{equation}\label{eq:torque}
	T_i:=\epsilon_{ijk}\int_{\partial S} \text{d}C'\ f_k(\mathbf{r'})r_j'={}-\epsilon_{ijk}T_{jk},
\end{equation}
where $\epsilon_{ijk}$ denotes the Levi-Civit\`{a} tensor.
We for this purpose amend the two-dimensional space by a third dimension associated with the direction $\hat{\mathbf{z}}$. The vectors of torque $\mathbf{T}$ and rotation $\mathbf{\Omega}$ (see below) point into that third direction. In this way, we guarantee that rotational displacements will be confined  to our two-dimensional plane. Simultaneously, we may use the convenient notation of the vector product to perform our evaluations.
Using
\begin{equation}
	T_{jk}\frac{\partial  G_{ij}(\mathbf{r})}{\partial r_k}={}-\frac{1}{2}\epsilon_{jkl}T_{l}\frac{\partial  G_{ij}(\mathbf{r})}{\partial r_k}=\frac{1}{2}(\mathbf{T}\times\nabla)_j G_{ij},
\end{equation} 
Eq.~(\ref{eq:u_series}) is rewritten as
\begin{equation}\label{eq:u_exp}
	\mathbf{u}(\mathbf{r})=\underline{\mathbf{G}}(\mathbf{r})\cdot\mathbf{F}-\left(\frac{1}{2}\mathbf{T\times\nabla+\underline{\mathbf{S}}\cdot\nabla}\right)\cdot\underline{\mathbf{G}}(\mathbf{r}).
\end{equation}

\section{Displacement field induced by a uniformly translated rigid circular inclusion}\label{sec:force}

Now we concentrate on one rigid disk-like inclusion of radius $a$, which is centered at position $\mathbf{r}_0$. Our goal is to find an analytical expression for the displacement field resulting when an external force $\mathbf{F}$ drags the inclusion. As mentioned above, we assume no-slip boundary conditions along the circumference, i.e., the surrounding medium sticks to $\partial S$. Therefore,
\begin{equation}\label{eq:boundary_trans}
	\mathbf{u}(\mathbf{r}\in\partial S)=\mathbf{U},
\end{equation}
where $\mathbf{U}$ represents the overall translation of the inclusion due to the external force. Thus, the force that pulls on the sphere is transmitted to the surrounding elastic medium, which leads to elastic distortions.

In the linear regime $\mathbf{F}\propto\mathbf{U}$. Therefore, we use the ansatz $\mathbf{u}(\mathbf{r})\propto\underline{\mathbf{G}}(\mathbf{r}-\mathbf{r}_0)\cdot\mathbf{F}\propto\underline{\mathbf{G}}(\mathbf{r}-\mathbf{r}_0)\cdot\mathbf{U}$. To satisfy the boundary condition Eq.~(\ref{eq:boundary_trans}), we can further demand the expression to be independent of $\mathbf{r}$ on $\partial S$. If we introduce an additional differential operator that acts on $\underline{\mathbf{G}}(\mathbf{r}-\mathbf{r}_0)$, we can satisfy all our requirements by setting
\begin{equation}
	\mathbf{u}(\mathbf{r}) = \frac{16\pi(1-\nu)\mu}{1-2(3-4\nu)\ln a}\left(1+\frac{a^2}{4}\nabla^2\right)\underline{\mathbf{G}}\left(\mathbf{r}-\mathbf{r}_0\right)\cdot \mathbf{U}.
\end{equation}
This expression uniquely solves Eq.~(\ref{eq:NCE}) and satisfies Eq.~(\ref{eq:boundary_trans}), because 
\begin{equation}
	\left.\left(1+\frac{a^2}{4}\nabla^2\right)\underline{\mathbf{G}}\left(\mathbf{r}-\mathbf{r}_0\right)\right|_{|\mathbf{r}-\mathbf{r}_0|=a} = \frac{1-2(3-4\nu)\ln a}{16\pi(1-\nu)\mu} \hat{\underline{\mathbf{I}}}.
\end{equation}

In the spirit of Eq.~(\ref{eq:uGF}), particularly for small $a$ and large $|\mathbf{r}-\mathbf{r}_0|>a$, we identify
\begin{equation}
	\mathbf{F} = \frac{16\pi(1-\nu)\mu}{1-2(3-4\nu)\ln a}\mathbf{U},
\end{equation}
or, analogously,
\begin{equation}\label{eq:u_trans_dS}
	\mathbf{u}(\mathbf{r}\in \partial S) = \mathbf{U} = \frac{1-2(3-4\nu)\ln a}{16\pi(1-\nu)\mu}\mathbf{F}.
\end{equation}
Thus, we may rewrite the displacement field as
\begin{equation}\label{eq:u_trans}
	\mathbf{u}(\mathbf{r}) = \left(1+\frac{a^2}{4}\nabla^2\right)\underline{\mathbf{G}}\left(\mathbf{r}-\mathbf{r}_0\right)\cdot \mathbf{F}.
\end{equation}
For later reference, we insert the result into Eq.~(\ref{eq:uGf_int}) to find for a disk-shaped inclusion and $|\mathbf{r}-\mathbf{r}_0|>a$
\begin{equation}\label{eq:Gf_int}
	\int_{\partial S} \underline{\mathbf{G}}\left(\mathbf{r}-\mathbf{r'}\right)\cdot\mathbf{f}(\mathbf{r'})\ \text dC'= \left(1+\frac{a^2}{4}\nabla^2\right)\underline{\mathbf{G}}\left(\mathbf{r}-\mathbf{r}_0\right)\cdot \mathbf{F}.
\end{equation}

\section{Displacement field induced by a uniformly rotated rigid circular inclusion}\label{sec:torque}

We keep the setting of Sec.~\ref{sec:force}, but consider a torque $\mathbf{T}$ that is exerted on the inclusion instead of the force $\mathbf{F}$. The resulting rotation is described by a rotation vector $\mathbf{\Omega}$
and leads to the boundary condition
\begin{equation}\label{eq:boundary_rot}
	\mathbf{u}(\mathbf{r}\in\partial S)=\mathbf{\Omega}\times\left(\mathbf{r}-\mathbf{r}_0\right)
\end{equation}
on $\partial S$. To physically remain with our system in a two-dimensional setting, we must restrict the torque $\mathbf{T}$ and thus the rotation vector $\mathbf{\Omega}$ to be perpendicular to the material plane.

From the second term on the right-hand side of Eq.~(\ref{eq:u_exp}) we find the resulting displacement field
\begin{equation}\label{eq:u_rot}
	\mathbf{u}(\mathbf{r})=\left(\frac{a}{|\mathbf{r}-\mathbf{r}_0|}\right)^2\mathbf{\Omega}\times\left(\mathbf{r}-\mathbf{r}_0\right),
\end{equation}
where 
\begin{equation}\label{eq:T_Omega}
	\mathbf{T} = 4\pi\mu a^2 \mathbf{\Omega}.
\end{equation}
This solution satisfies Eq.~(\ref{eq:NCE}) as well as the boundary condition Eq.~(\ref{eq:boundary_rot}).

\section{Faxén's laws}\label{sec:Faxen}

The next question that arises is how the inclusion reacts when exposed to a displacement field $\mathbf{u}(\mathbf{r})$ induced by other sources in the surrounding medium. Nevertheless, we still allow the inclusion to be subject to an imposed external force $\mathbf{F}$ or torque $\mathbf{T}$. The overall force line density $\mathbf{f}(\mathbf{r})$, see Eq.~(\ref{eq:uGf_int}), along the circumference $\partial S$ resulting from both contributions in combination with the rigidity of the inclusion is transmitted to the surrounding medium. All resulting displacements can simply be superimposed, because Eq.~(\ref{eq:NCE}) is linear. This leads to the stick boundary condition 
\begin{equation}\label{eq:boundary_Faxen}
	U_i+[\mathbf{\Omega}\times(\mathbf{r}-\mathbf{r}_0)]_i
	= \int_{\partial S} G_{ij}(\mathbf{r}-\mathbf{r'})f_j(\mathbf{r'})\text dC' + u_i(\mathbf{r})
\end{equation}
for $\mathbf{r}\in\partial S$.

The left-hand side of this equation describes the displacement of each surface point of the inclusion due to  the rigid translation $\mathbf{U}$ and rotation $\mathbf{\Omega}$. Conversely, the right-hand side of the equation quantifies the displacement of each element of the surrounding medium anchored to the surface of the inclusion. Here, the first term results from the circumference force line density exerted by the inclusion. The second term is the externally imposed displacement field. To satisfy the stick boundary condition, the inclusion must displace to the same amount as the elastic medium at each position $\mathbf{r}\in \partial S$ setting left- and right-hand sides of Eq.~(\ref{eq:boundary_Faxen}) equal.

First we focus on the translation of the inclusion. For this purpose, we integrate Eq.~(\ref{eq:boundary_Faxen}) over $\partial S$, which results in
\begin{equation}\label{eq:boundary_Faxen_trans}
	2\pi a U_i = \int_{\partial S} \int_{\partial S} G_{ij}(\mathbf{r}-\mathbf{r'})f_j(\mathbf{r'})\text dC' \text dC + \int_{\partial S}u_i(\mathbf{r}) \text dC .
\end{equation}
The first term on the right-hand side is evaluated using Eqs.~(\ref{eq:u_trans_dS})--(\ref{eq:Gf_int}) in inverse order.

To evaluate the second term on the right-hand side, we expand $u_i(\mathbf{r})$ around $\mathbf{r}=\mathbf{r}_0$ as 
\begin{align}\label{eq:u_Taylor}
	&u_i(\mathbf{r}) \nonumber\\&= \ u_i(\mathbf{r}_0) + (\mathbf{r}-\mathbf{r}_0)_j[\nabla_j u_i(\mathbf{r})]_{\mathbf{r}=\mathbf{r}_0} \nonumber\\ 
	&\quad + \frac{1}{2}(\mathbf{r}-\mathbf{r}_0)_j(\mathbf{r}-\mathbf{r}_0)_k[\nabla_j \nabla_k u_i(\mathbf{r})]_{\mathbf{r}=\mathbf{r}_0} \nonumber\\
	&\quad+\frac{1}{3!}(\mathbf{r}-\mathbf{r}_0)_j(\mathbf{r}-\mathbf{r}_0)_k(\mathbf{r}-\mathbf{r}_0)_l[\nabla_j \nabla_k \nabla_l u_i(\mathbf{r})]_{\mathbf{r}=\mathbf{r}_0} \nonumber \\
	&\quad + ... .
\end{align}
During the integration in Eq.~(\ref{eq:boundary_Faxen_trans}), the terms odd in $(\mathbf{r}-\mathbf{r}_0)$ vanish because of symmetry. Moreover, from Eq.~(\ref{eq:NCE}), we find that $\nabla^4\mathbf{u}(\mathbf{r})=\mathbf{0}$. Together with
\begin{equation}\label{eq:int_rr}
	\int_{\partial S} r_j r_k dC = \pi a^3\delta_{jk},
\end{equation}
the second term on the right-hand side of Eq.~(\ref{eq:boundary_Faxen_trans}) becomes
\begin{align}
	&\int_{\partial S}u_i(\mathbf{r}) \text dC \nonumber\\
	&= 2\pi a u_i(\mathbf{r}_0) \nonumber\\
	&\quad + \frac{1}{2} \int_{\partial S}(\mathbf{r}-\mathbf{r}_0)_j(\mathbf{r}-\mathbf{r}_0)_k[\nabla_j \nabla_k u_i(\mathbf{r})]_{\mathbf{r}=\mathbf{r}_0}  \text dC \nonumber\\
	&= 2 \pi a \left.\left(1+\frac{a^2}{4}\nabla^2\right)u_i(\mathbf{r})\right|_{\mathbf{r}=\mathbf{r}_0}.
\end{align}
Together, we find from Eq.~(\ref{eq:boundary_Faxen_trans})
\begin{equation}\label{eq:Faxen_trans_F}
	\mathbf{U} = \frac{1-2(3-4\nu)\ln a}{16\pi(1-\nu)\mu}\mathbf{F}+\left.\left(1+\frac{a^2}{4}\nabla^2\right)\mathbf{u}(\mathbf{r})\right|_{\mathbf{r}=\mathbf{r}_0}.
\end{equation}
In this expression, the first term on the right-hand side recovers Eq.~(\ref{eq:u_trans_dS}) and therefore directly results from the imposed force $\mathbf{F}$. Thus the remaining part on the right-hand side of Eq.~(\ref{eq:Faxen_trans_F}) originates from the imposed displacement field $\mathbf{u}(\mathbf{r})$. 
Thus, we obtain the first Faxén law
\begin{equation}\label{eq:Faxen_1}
	\mathbf{U}^{\text{Faxén}} = \left.\left(1+\frac{a^2}{4}\nabla^2\right)\mathbf{u}(\mathbf{r})\right|_{\mathbf{r}=\mathbf{r}_0},
\end{equation}
which describes the displacement of a rigid disk-like inclusion solely due to the displacement of the surrounding medium.

Next, we focus on the rotation vector and the stresslet.  
To this end, we multiply Eq.~(\ref{eq:boundary_Faxen}) by $(\mathbf{r}-\mathbf{r}_0)_k$ and then integrate over $\partial S$,
\begin{align}\label{eq:boundary_Faxen_rot}
	\int_{\partial S}& (\mathbf{r}-\mathbf{r}_0)_k [\mathbf{\Omega}\times(\mathbf{r}-\mathbf{r}_0)]_i \text dC \nonumber \\
	 =& \int_{\partial S} \int_{\partial S} (\mathbf{r}-\mathbf{r}_0)_k G_{ij}(\mathbf{r}-\mathbf{r'})f_j(\mathbf{r'})\text dC \text dC' \nonumber \\	
	 &+ \int_{\partial S} (\mathbf{r}-\mathbf{r}_0)_k u_i(\mathbf{r}) \text dC.
\end{align}

The integral on the left-hand side is directly calculated via Eq.~(\ref{eq:int_rr}), leading to
\begin{equation}\label{eq:Faxe_rot_res1}
	\int_{\partial S} (\mathbf{r}-\mathbf{r}_0)_k [\mathbf{\Omega}\times(\mathbf{r}-\mathbf{r}_0)]_i\ \text dC = \pi a^3\epsilon_{izk}\Omega_z,
\end{equation}
where $\Omega_z=\hat{\mathbf{z}}\cdot\mathbf{\Omega}$.
On the right-hand side of Eq.~(\ref{eq:boundary_Faxen_rot}), we first concentrate on the inner integral of the first term. To calculate it, we set $\mathbf{r}''=\mathbf{r}-\mathbf{r}_0$ and express the Green's function through its Fourier transform,
\begin{align}\label{eq:int_Fourier}
	\int_{\partial S}& (\mathbf{r}-\mathbf{r}_0)_k G_{ij}(\mathbf{r}-\mathbf{r'})\text dC \nonumber \\
	=& \int_{\partial S} G_{ij}(\mathbf{r''}-\mathbf{r'}+\mathbf{r}_0) r''_k\text dC'' \nonumber \\
	=& \frac{1}{(2\pi)^2} \int_{\partial S} \text dC'' \ \int \text d^2k\   \tilde{G}_{ij}(\mathbf{k}) r''_k e^{i\mathbf{k}\cdot(\mathbf{r''}-\mathbf{r'}+\mathbf{r}_0)}.
\end{align}
The integral over $\text dC''$ can be calculated as
\begin{align}\label{eq:int_r''}
	\int_{\partial S} \text dC'' \ r''_k e^{i\mathbf{k}\cdot\mathbf{r''}} 
	&={}-i\nabla_{\mathbf{k},k} \int_{\partial S} \text dC''\  e^{i\mathbf{k}\cdot\mathbf{r''}} \nonumber \\
	&={}-2\pi i a \hat{k}_k \frac{\text d}{\text d k} J_0 (ka) \nonumber \\
	&= 2\pi i a^2 \hat{k}_k J_1 (ka),
\end{align}
where $J_0$ and $J_1$ are Bessel functions of the first kind.
Inserting Eq.~(\ref{eq:uGF}) into Eq.~(\ref{eq:NCE}) for $\mathbf{f}_{\mathrm{b}}(\mathbf{r})=\mathbf{F}_0\delta(\mathbf{r}-\mathbf{r}_0)$ and Fourier transforming the whole equation, we obtain the components of the Fourier-transform of the Green's function as
\begin{equation}\label{eq:G_Fourier}
	\tilde{G}_{ij}=\frac{1}{\mu k^2}\left[\delta_{ij}-\frac{1}{2(1-\nu)}\hat{k}_i\hat{k}_j\right].
\end{equation}

\begin{widetext}
Next we insert Eqs.~(\ref{eq:int_r''}) and (\ref{eq:G_Fourier}) into Eq.~(\ref{eq:int_Fourier}), which leads us to 
\begin{align}
	&\frac{1}{(2\pi)^2} \int_{\partial S} \text dC'' \ \int \text d^2k\   \tilde{G}_{ij}(\mathbf{k}) r''_k e^{i\mathbf{k}\cdot(\mathbf{r''}-\mathbf{r'}+\mathbf{r}_0)} \nonumber\\
	&=\frac{i a^2}{2\pi \mu} \int_0^{2\pi} \text d\varphi_{\mathbf{k}}\  \left[\delta_{ij}-\frac{1}{2(1-\nu)}\hat{k}_i\hat{k}_j\right] \hat{k}_k \int_0^{\infty} \text dk\ \frac{1}{k} e^{-i\mathbf{k}\cdot(\mathbf{r'}-\mathbf{r}_0)} J_1 (ka).
\end{align}
Here, we split $\int \text d^2k$ into $\int_0^{2\pi}\text d\varphi_{\mathbf{k}}\int_0^\infty k\ \text dk$. With the help of \textit{Mathematica} \cite{Mathematica}, the $\text dk$-integral in the relevant range of $\mathbf{r}'$ can be evaluated to
\begin{equation}
	\int_0^{\infty} \text dk\ \frac{J_1 (ka)}{k} e^{-ik\hat{\mathbf{k}}\cdot(\mathbf{r'}-\mathbf{r}_0)} = \sqrt{1-\left(\frac{\hat{\mathbf{k}}\cdot(\mathbf{r'}-\mathbf{r}_0)}{a}\right)^2}-i\frac{\hat{\mathbf{k}}\cdot(\mathbf{r'}-\mathbf{r}_0)}{a} \quad\quad\text{for}\quad -1< \frac{\hat{\mathbf{k}}\cdot(\mathbf{r'}-\mathbf{r}_0)}{a}<1.
\end{equation}
The remaining integral can be calculated using
\begin{equation}
	\int_0^{2\pi} \text d\varphi_{\mathbf{k}}\  \hat{k}_k\hat{k}_l = \delta_{kl}\pi
\end{equation}
and
\begin{equation}\label{eq:int_kkkk}
	\int_0^{2\pi} \text d\varphi_{\mathbf{k}}\  \hat{k}_i\hat{k}_j\hat{k}_k\hat{k}_l
	=\frac{\pi}{4}(\delta_{ij}\delta_{kl}+\delta_{ik}\delta_{jl}+\delta_{il}\delta_{jk}).
\end{equation}
It results in
\begin{align}\label{eq:Faxe_rot_res2}
	&\int_{\partial S} \left(\mathbf{r}-\mathbf{r}_0\right)_k G_{ij}\left(\mathbf{r}-\mathbf{r'}\right)f_j\left(\mathbf{r'}\right)\text dC \nonumber\\
	&=\frac{i a^2}{2\pi \mu} \int_0^{2\pi} \text d\varphi_{\mathbf{k}}\  \left[\delta_{ij}-\frac{1}{2(1-\nu)}\hat{k}_i\hat{k}_j\right] \hat{k}_k \left(\sqrt{1-\frac{1}{a^2}\left(\hat{k}_l\left(\mathbf{r'}-\mathbf{r}_0\right)_l\right)^2}-\frac{i}{a}\hat{k}_l\left(\mathbf{r'}-\mathbf{r}_0\right)_l\right)\nonumber\\
	&= \frac{a}{8\mu} \left\{4\left(\mathbf{r'}-\mathbf{r}_0\right)_k f_i\left(\mathbf{r}'\right)-\frac{1}{2(1-\nu)}\big[\left(\mathbf{r'}-\mathbf{r}_0\right)_k f_i\left(\mathbf{r}'\right) + \left(\mathbf{r'}-\mathbf{r}_0\right)_i f_k \left(\mathbf{r}'\right)+ \left(\mathbf{r'}-\mathbf{r}_0\right)_l f_l \left(\mathbf{r}'\right)\delta_{ik} \big]\right\}.
\end{align}

To calculate the second integral on the right-hand side of Eq.~(\ref{eq:boundary_Faxen_rot}), we insert again the Taylor expansion of $\mathbf{u}(\mathbf{r})$ from Eq.~(\ref{eq:u_Taylor}). Using Eqs.~(\ref{eq:int_rr}) and (\ref{eq:int_kkkk}) for $(\mathbf{r}-\mathbf{r}_0)$ instead of $\hat{\mathbf{k}}$ we evaluate this integral to
\begin{align}\label{eq:Faxe_rot_res3}
	\int_{\partial S} (\mathbf{r}-\mathbf{r}_0)_k u_i(\mathbf{r}) \text dC 
	&= \int_{\partial S} (\mathbf{r}-\mathbf{r}_0)_k (\mathbf{r}-\mathbf{r}_0)_j[\nabla_j u_i(\mathbf{r})]_{\mathbf{r}=\mathbf{r}_0} \text dC \nonumber\\
	&\quad + \frac{1}{6} \int_{\partial S} (\mathbf{r}-\mathbf{r}_0)_k (\mathbf{r}-\mathbf{r}_0)_j(\mathbf{r}-\mathbf{r}_0)_l(\mathbf{r}-\mathbf{r}_0)_m[\nabla_j \nabla_l \nabla_m u_i(\mathbf{r})]_{\mathbf{r}=\mathbf{r}_0}\text dC \nonumber\\
	&=\pi a^3 \delta_{jk} \nabla_j u_i(\mathbf{r})|_{\mathbf{r}=\mathbf{r}_0}
		+\frac{1}{6}\frac{\pi a^5}{4}(\delta_{jk}\delta_{lm}+\delta_{jl}\delta_{km}+\delta_{jm}\delta_{kl})[\nabla_j \nabla_l \nabla_m u_i(\mathbf{r})]_{\mathbf{r}=\mathbf{r}_0} \nonumber\\
	&= \pi a^3 \left.\left(1+\frac{a^2}{8}\nabla^2\right)\nabla_k u_i(\mathbf{r})\right|_{\mathbf{r}=\mathbf{r}_0}.
\end{align}
Combining the above results in Eqs.~(\ref{eq:boundary_Faxen_rot}), (\ref{eq:Faxe_rot_res1}), (\ref{eq:Faxe_rot_res2}), and (\ref{eq:Faxe_rot_res3}) leads us to
\begin{align}\label{eq:Faxen_rot}
	\pi a^3\epsilon_{izk}\Omega_z =& \frac{a}{8\mu} \int_{\partial S} \text dC'\ \left\{4(\mathbf{r'}-\mathbf{r}_0)_k f_i(\mathbf{r}')-\frac{1}{2(1-\nu)}\big[\left(\mathbf{r'}-\mathbf{r}_0\right)_k f_i\left(\mathbf{r}'\right) + \left(\mathbf{r'}-\mathbf{r}_0\right)_i f_k \left(\mathbf{r}'\right)+ \left(\mathbf{r'}-\mathbf{r}_0\right)_l f_l\left(\mathbf{r}'\right) \delta_{ik} \big]\right\} \nonumber \\
	&+\pi a^3 \left.\left(1+\frac{a^2}{8}\nabla^2\right)\nabla_k u_i(\mathbf{r})\right|_{\mathbf{r}=\mathbf{r}_0}.
\end{align}

We split this equation into a symmetric and an antisymmetric part. To obtain the antisymmetric part, we multiply Eq.~(\ref{eq:Faxen_rot}) by $\epsilon_{izk}=\epsilon_{zki}$. Using $\epsilon_{izk}\epsilon_{izk}=\delta_{zz}\delta_{kk}-\delta_{zk}\delta_{zk}=2$, this leads to
\begin{align}\label{eq:Omega_z}
	2\pi a^3\Omega_z =& \frac{a}{8\mu} \int_{\partial S} \text dC'\  \epsilon_{zki}\left\{4(\mathbf{r'}-\mathbf{r}_0)_k f_i(\mathbf{r}')-\frac{1}{2(1-\nu)}\big[\left(\mathbf{r'}-\mathbf{r}_0\right)_k f_i \left(\mathbf{r}'\right)+ \left(\mathbf{r'}-\mathbf{r}_0\right)_i f_k \left(\mathbf{r}'\right)+ \left(\mathbf{r'}-\mathbf{r}_0\right)_l f_l\left(\mathbf{r}'\right) \delta_{ik} \big]\right\} \nonumber \\
	&+\pi a^3 \left.\left(1+\frac{a^2}{8}\nabla^2\right)\epsilon_{zki}\nabla_k u_i(\mathbf{r})\right|_{\mathbf{r}=\mathbf{r}_0}.
\end{align}
From Eq.~(\ref{eq:NCE}), we infer that $\nabla\times\nabla^2\mathbf{u}(\mathbf{r}=\mathbf{r}_0)=\mathbf{0}$, which reduces the last term in Eq.~(\ref{eq:Omega_z}). Overall, we obtain from Eq.~(\ref{eq:Omega_z})
\begin{equation}
	\Omega_z = \frac{1}{4 \pi \mu a^2} \int_{\partial S} \text dC'\  \epsilon_{zki}(\mathbf{r'}-\mathbf{r}_0)_k f_i+\frac{1}{2} \left. \epsilon_{zki}\nabla_k u_i(\mathbf{r})\right|_{\mathbf{r}=\mathbf{r}_0}.
\end{equation}
Using Eq.~(\ref{eq:torque}), we find
\begin{equation}\label{eq:Faxen_rot_T}
	\mathbf{\Omega} = \frac{1}{4 \pi \mu a^2} \mathbf{T} + \left.\frac{1}{2}\big(\nabla\times \mathbf{u}\left(\mathbf{r}\right)\big)\right|_{\mathbf{r}=\mathbf{r}_0}.
\end{equation}
In analogy to the translational case in Eq.~(\ref{eq:Faxen_trans_F}), the first term on the right-hand side recovers Eq.~(\ref{eq:T_Omega}) and therefore describes the rotation of the inclusion due to the external torque $\mathbf{T}$. Conversely, the second term on the right-hand side of Eq.~(\ref{eq:Faxen_rot_T}) arises solely from the displacement field in the surrounding material. Thus, we obtain the second Fax\'{e}n law in the form
\begin{equation}\label{eq:Faxen_2}
	\mathbf{\Omega}^{\text{Faxén}} =  \left.\frac{1}{2} \big(\nabla\times \mathbf{u}\left(\mathbf{r}\right)\big)\right|_{\mathbf{r}=\mathbf{r}_0}.
\end{equation}
We recall that the displacement field $\mathbf{u}(\mathbf{r})$ is confined to the two-dimensional plane of our system. Therefore, both $\mathbf{\Omega}$ and $\mathbf{T}$ consistently point into the corresponding normal direction. In this way, they in turn only induce rotational displacements within the two-dimensional plane.

Finally, we list the symmetric part of Eq.~(\ref{eq:Faxen_rot}). Its left-hand side is antisymmetric so that it does not contribute, and we obtain
\begin{align}\label{eq:Faxen_stress}
	0 &= \frac{a}{8\mu} \frac{1}{2(1-\nu)} \int_{\partial S} \text dC'\ \left\{(3-4\nu)\big[\left(\mathbf{r'}-\mathbf{r}_0\right)_i f_k\left(\mathbf{r}'\right)+\left(\mathbf{r'}-\mathbf{r}_0\right)_k f_i\left(\mathbf{r}'\right)\big]-\left(\mathbf{r'}-\mathbf{r}_0\right)_j f_j\left(\mathbf{r}'\right) \delta_{ik}\right\} \nonumber \\
	&\quad +\pi a^3 \left.\left(1+\frac{a^2}{8}\nabla^2\right)\frac{1}{2}\big[\nabla_i u_k\left(\mathbf{r}\right)+\nabla_k u_i\left(\mathbf{r}\right)\big]\right|_{\mathbf{r}=\mathbf{r}_0} \nonumber\\
	&:=\frac{1}{2}(A_{ik}+A_{ki}).
\end{align}
Since, obviously from this equation, the trace $A_{jj}$ vanishes, we may add it to Eq.~(\ref{eq:Faxen_stress}) in the form
\begin{equation}
	\frac{1}{4(1-2\nu)}A_{jj}\delta_{ik} 
	= \frac{a}{8\mu} \frac{1}{2(1-\nu)} \int_{\partial S} \text dC'\ (\mathbf{r'}-\mathbf{r}_0)_j f_j(\mathbf{r}') \delta_{ik}
	+\frac{\pi a^3}{4(1-2\nu)} \left.\left(1+\frac{a^2}{8}\nabla^2\right)\nabla_j u_j(\mathbf{r})\delta_{ik}\right|_{\mathbf{r}=\mathbf{r}_0}.
\end{equation}
As a result, we find from Eq.~(\ref{eq:Faxen_stress})
\begin{align}
	0&=\frac{1}{2}(A_{ik}+A_{ki})+\frac{1}{4(1-2\nu)}A_{jj}\delta_{ik} \nonumber\\
	&= \frac{a}{8\mu} \frac{1}{2(1-\nu)} \int_{\partial S} \text dC'\ \big\{(3-4\nu)\big[\left(\mathbf{r'}-\mathbf{r}_0\right)_i f_k\left(\mathbf{r}'\right)+\left(\mathbf{r'}-\mathbf{r}_0\right)_k f_i\left(\mathbf{r}'\right)\big]\big\} \nonumber \\
	&\quad+\pi a^3 \left.\left(1+\frac{a^2}{8}\nabla^2\right)\frac{1}{2}\big[\nabla_i u_k\left(\mathbf{r}\right)+\nabla_k u_i\left(\mathbf{r}\right)\big]\right|_{\mathbf{r}=\mathbf{r}_0} +\frac{\pi a^3}{4(1-2\nu)} \left.\left(1+\frac{a^2}{8}\nabla^2\right)\nabla_j u_j(\mathbf{r})\delta_{ik}\right|_{\mathbf{r}=\mathbf{r}_0} \nonumber\\
	&= \frac{a}{8\mu} \frac{(3-4\nu)}{(1-\nu)} S_{ik} 
	+\frac{\pi a^3}{4} \left(1+\frac{a^2}{8}\nabla^2\right)\left.\left(2\big[\nabla_i u_k(\mathbf{r})+\nabla_k u_i(\mathbf{r})\big] 
	+\frac{1}{(1-2\nu)} \nabla_j u_j(\mathbf{r})\delta_{ik}\right)\right|_{\mathbf{r}=\mathbf{r}_0},
\end{align}
where we have used the definition of the components of the stresslet $S_{ik}$ according to Eq.~(\ref{eq:stresslet}).
From here, we obtain
\begin{equation}\label{eq:Faxen_S}
	\underline{\mathbf{S}}={}-\frac{2 \pi(1-\nu)\mu a^2}{(3-4\nu)} \left(1+\frac{a^2}{8}\nabla^2\right)\left.\left(2\left\{\nabla \mathbf{u}(\mathbf{r})+\left[\nabla \mathbf{u}(\mathbf{r})\right]^{T}\right\} 
		+\frac{1}{(1-2\nu)} \hat{\underline{\mathbf{I}}} \nabla\cdot \mathbf{ u}(\mathbf{r})\right)\right|_{\mathbf{r}=\mathbf{r}_0},
\end{equation}
where $[\ ]^T$ marks the transpose. This expression quantifies the stress that a rigid circular inclusion exerts on its surrounding due to its resistance against deformation, if it is exposed to a displacement field $\mathbf{u}(\mathbf{r})$ in the elastic medium. Conversely, the stresslet that the surrounding medium exerts on the rigid inclusion follows as
\begin{equation}
	\underline{\mathbf{S}}^{\text{Faxén}}={}-\underline{\mathbf{S}}.
\end{equation}
\end{widetext}

\section{Displaceability and rotateability matrices}\label{sec:matrix}
Now we know how each inclusion reacts to forces, torques, and imposed displacement fields. Next, we address the coupling of $N$ identical inclusions embedded in the elastic surrounding medium through deformations of this medium. We suppose that every inclusion $j$ may be subject to a force $\mathbf{F}_j$ and torque $\mathbf{T}_j$ imposed from outside, i.e., not resulting from the action of the elastic medium on the inclusion. These forces and torques directly lead to translations $\mathbf{U}_j$ and rotations $\mathbf{\Omega}_j$ of the inclusion. In turn, these reconfigurations imply displacements of the surrounding medium quantified by a corresponding displacement field. All other inclusions feel this displacement field. The inclusions counteract displacement fields that would imply their deformation, because of their rigidity. Corresponding counterstresses induce additional displacements in the surrounding medium, which in turn affect the inclusions.
Starting from the external forces $\mathbf{F}_j$ and torques $\mathbf{T}_j$, $j=1,...,N$, we calculate the displacements $\mathbf{U}_j$ and rotations $\mathbf{\Omega}_j$ resulting from these effects in the form
\begin{equation} \label{eq:matrix}
	\left(\begin{array}{@{}c@{}} 
		\mathbf{U}_1\\
		\vdots\\
		\mathbf{U}_N\vspace{0.2cm}\\	
		\mathbf{\Omega}_1\\
		\vdots\\
		\mathbf{\Omega}_N 
	\end{array}\right)
	=
	\left(\begin{array}{@{}c @{} c @{} c @{} c @{} c @{} c @{}}
		\underline{\mathbf{M}}^{\mathrm{tt}}_{11} & \cdots & \underline{\mathbf{M}}^{\mathrm{tt}}_{1N} & \underline{\mathbf{M}}^{\mathrm{tr}}_{11} & \cdots & \underline{\mathbf{M}}^{\mathrm{tr}}_{1N}\\
		\vdots & \ddots & \vdots & \vdots & \ddots & \vdots\\
		\underline{\mathbf{M}}^{\mathrm{tt}}_{N1} & \cdots & \underline{\mathbf{M}}^{\mathrm{tt}}_{NN} & \underline{\mathbf{M}}^{\mathrm{tr}}_{N1} & \cdots & \underline{\mathbf{M}}^{\mathrm{tr}}_{NN}\vspace{0.2cm}\\
		\underline{\mathbf{M}}^{\mathrm{rt}}_{11} & \cdots & \underline{\mathbf{M}}^{\mathrm{rt}}_{1N} & \underline{\mathbf{M}}^{\mathrm{rr}}_{11} & \cdots & \underline{\mathbf{M}}^{\mathrm{rr}}_{1N}\\
		\vdots & \ddots & \vdots & \vdots & \ddots & \vdots\\
		\underline{\mathbf{M}}^{\mathrm{rt}}_{N1} & \cdots & \underline{\mathbf{M}}^{\mathrm{rt}}_{NN} & \underline{\mathbf{M}}^{\mathrm{rr}}_{N1} & \cdots & \underline{\mathbf{M}}^{\mathrm{rr}}_{NN}\\
	\end{array}\right)
	\cdot
	\left(\begin{array}{@{}c@{}}
		\mathbf{F}_1\\
		\vdots\\
		\mathbf{F}_N\vspace{0.2cm}\\	
		\mathbf{T}_1\\
		\vdots\\
		\mathbf{T}_N 
	\end{array}\right).
\end{equation}
The matrix on the right-hand side contains four kinds of submatrices. First, there are the $\underline{\mathbf{M}}^{\mathrm{tt}}_{ij}$-matrices, $i,j=1,...,N$, which describe translation--translation couplings (translations due to forces). The second kind is given by the $\underline{\mathbf{M}}^{\mathrm{tr}}_{ij}$-matrices, which derive from the translation--rotation couplings (translations due to torques). The next ones are the $\underline{\mathbf{M}}^{\mathrm{rt}}_{ij}$-matrices, which describe how the inclusions rotate in response to imposed forces (rotation--translation couplings). The last kind is represented by the $\underline{\mathbf{M}}^{\mathrm{rr}}_{ij}$-matrices, which give the rotation from the applied torques (rotation--rotation couplings).
In Secs.~\ref{sec:interaction_two} and \ref{sec:interaction_three}, we explicitly calculate these matrices up to (including) the third order in inverse distances between the inclusions. The natural couplings between the inclusions mediated by the elastic environment are taken into account by these matrices.

\section{Two-body interactions}\label{sec:interaction_two}
In this section, we explicitly calculate expressions for the submatrices introduced in Eq.~(\ref{eq:matrix}). Starting from the applied forces and torques acting on each inclusion, we evaluate how all other inclusions react to the displacement fields induced in this way. We follow an iterative scheme, termed the method of reflections \cite{dhont1996introduction}. To lowest order, displacement fields are introduced into the system by the direct response of individual inclusions to forces and torques that they are exposed to, as if the other inclusions were absent. Then the response of all inclusions to the displacement fields induced in this way are evaluated. Counterstresses emerge because of the rigidity of the inclusions and their resistance to deformations, which leads to additional displacement fields. At the end, because of the linearity of the Navier--Cauchy equations, the different contributions are simply superimposed.

\subsection{Forces imposed on or induced between the inclusions}\label{sec:interaction_two_force}
First, we consider two identical inclusions $i$ and $j$ of radius $a$ at different positions $\mathbf{r}_i$ and $\mathbf{r}_j$. They are subject to the forces $\mathbf{F}_i$ and $\mathbf{F}_j$, respectively.
To lowest order, each inclusion directly reacts to the force acting on it as if the other inclusions were not present. This leads to their displacements in analogy to Eq.~(\ref{eq:u_trans_dS}),
\begin{equation}\label{eq:Ui(0)_force}
	\mathbf{U}_i^{(0)} = \mathbf{u}^{(0)}(\mathbf{r}\in \partial S_i) =  \frac{1-2(3-4\nu)\ln a}{16\pi(1-\nu)\mu}\mathbf{F}_i
\end{equation}
and
\begin{equation}
	\mathbf{U}_j^{(0)} = \mathbf{u}^{(0)}(\mathbf{r}\in \partial S_j) = \frac{1-2(3-4\nu)\ln a}{16\pi(1-\nu)\mu}\mathbf{F}_j.
\end{equation}
To this order, the displacement fields induced around the inclusions according to Eq.~(\ref{eq:u_trans}) are given by
\begin{equation}\label{eq:ui(0)}
	\mathbf{u}_i^{(0)}(\mathbf{r}) = \left(1+\frac{a^2}{4}\nabla^2\right)\underline{\mathbf{G}}\left(\mathbf{r}-\mathbf{r}_i\right)\cdot \mathbf{F}_i
\end{equation}
and
\begin{equation}\label{eq:uj(0)}
	\mathbf{u}_j^{(0)}(\mathbf{r}) = \left(1+\frac{a^2}{4}\nabla^2\right)\underline{\mathbf{G}}\left(\mathbf{r}-\mathbf{r}_j\right)\cdot \mathbf{F}_j,
\end{equation}
where the positions $\mathbf{r}_i$ and $\mathbf{r}_j$ of the inclusions enter.

To next order, the two inclusions affect each other through the displacement field that they induce in the surrounding medium. Inclusion $i$ is exposed to the displacement field $\mathbf{u}_j^{(0)}(\mathbf{r})$, which leads via Eq.~(\ref{eq:Faxen_1}) to the translation 
\begin{align}\label{eq:Ui(1)_force}
	\mathbf{U}_i^{(1)} &= \left.\left(1+\frac{a^2}{4}\nabla^2\right)\mathbf{u}_j^{(0)}(\mathbf{r})\right|_{\mathbf{r}=\mathbf{r}_i}\nonumber\\
	&= \frac{1}{8\pi(1-\nu)\mu}\left\{\left[-(3-4\nu)\ln r_{ij} +\left(\frac{a}{r_{ij}}\right)^2\right]\hat{\underline{\mathbf{I}}}\right.\nonumber\\
	&\quad\left.+\left[1-2\left(\frac{a}{r_{ij}}\right)^2\right]\hat{\mathbf{r}}_{ij}\hat{\mathbf{r}}_{ij}\right\}\cdot \mathbf{F}_j
\end{align}
and via Eq.~(\ref{eq:Faxen_2}) to the rotation
\begin{align}\label{eq:Omegai(1)_force}
	\mathbf{\Omega}_i^{(1)} &=  \frac{1}{2} (\nabla\times \mathbf{u}_j^{(0)}(\mathbf{r}))|_{\mathbf{r}=\mathbf{r}_i}=-\frac{1}{4\pi\mu r_{ij}} (\hat{\mathbf{r}}_{ij}\times\mathbf{F}_j).
\end{align}
Here, we defined $\mathbf{r}_{ij}=\mathbf{r}_i-\mathbf{r}_j$, $r_{ij}=|\mathbf{r}_{ij}|$, and $\hat{\mathbf{r}}_{ij}=\mathbf{r}_{ij}/r_{ij}$.
Apart from the inducing the translation and rotation of inclusion $i$, the displacement field $\mathbf{u}_j^{(0)}(\mathbf{r})$ would in general also deform it. Due to its rigidity, however, the inclusion resists its deformation and exerts the counterstress $\underline{\mathbf{S}}_i^{(1)}$ on the surrounding medium. These counterstresses add to the overall displacement field. The same happens for inclusion $j$, where corresponding expressions are obtained by exchanging indices $i$ and $j$. According to Eq.~(\ref{eq:u_exp}), the associated displacement fields are calculated via
\begin{equation}\label{eq:ui(1)}
	\mathbf{u}_i^{(1)}(\mathbf{r})={}-(\underline{\mathbf{S}}_i^{(1)}\cdot\nabla)\cdot\underline{\mathbf{G}}\left(\mathbf{r}-\mathbf{r}_i\right)
\end{equation}
and
\begin{equation}\label{eq:uj(1)}
	\mathbf{u}_j^{(1)}(\mathbf{r})={}-(\underline{\mathbf{S}}_j^{(1)}\cdot\nabla)\cdot\underline{\mathbf{G}}\left(\mathbf{r}-\mathbf{r}_j\right).
\end{equation}
They can simply be added to Eqs.~(\ref{eq:ui(0)}) and (\ref{eq:uj(0)}), because Eq.~(\ref{eq:NCE}) is linear. This requires to explicitly calculate the stresslets $\underline{\mathbf{S}}_i^{(1)}$ and $\underline{\mathbf{S}}_j^{(1)}$. For inclusion $j$, we obtain from Eqs.~(\ref{eq:Faxen_S}) and (\ref{eq:ui(0)})
\begin{align}\label{eq:Sj(1)_force}
	\underline{\mathbf{S}}_j^{(1)}&=-\frac{2 \pi(1-\nu)\mu a^2}{(3-4\nu)}\! \left(1+\frac{a^2}{8}\nabla^2\right)\!\! \left(\frac{1}{(1-2\nu)} \hat{\underline{\mathbf{I}}} \nabla\cdot \mathbf{u}_i^{(0)}(\mathbf{r})\right.\nonumber\\	
	&\quad\left.\left.+2\left\{\nabla \mathbf{u}_i^{(0)}(\mathbf{r})+\left[\nabla \mathbf{u}_i^{(0)}(\mathbf{r})\right]^{T}\right\} 
	\right)\right|_{\mathbf{r}=\mathbf{r}_j}\nonumber\\
	&= -\frac{1}{4(3-4\nu)}\frac{a^2}{r_{ij}}\Big\{4(1-2\nu)(\mathbf{F}_i\hat{\mathbf{r}}_{ij}+\hat{\mathbf{r}}_{ij}\mathbf{F}_i)\nonumber\\
	&\quad-2\hat{\underline{\mathbf{I}}}\hat{\mathbf{r}}_{ij}\cdot\mathbf{F}_i+8
	\hat{\mathbf{r}}_{ij}\hat{\mathbf{r}}_{ij}\hat{\mathbf{r}}_{ij}\cdot\mathbf{F}_i\Big\}+\mathcal{O}\big(r_{ij}^{-3}\big).
\end{align}
Using this expression, we can evaluate via Eq.~(\ref{eq:uj(1)}) the resulting translation of inclusion $i$ due to the disturbance  $\mathbf{u}_j^{(1)}(\mathbf{r})$. For this purpose, we use again the Faxén law Eq.~(\ref{eq:Faxen_1}), there inserting $\mathbf{u}_j^{(1)}(\mathbf{r})$ .
This leads to a translation
\begin{align} \label{eq:Ui(2)_force}
	\mathbf{U}_i^{(2)} &= \left.\left(1+\frac{a^2}{4}\nabla^2\right)\mathbf{u}_j^{(1)}(\mathbf{r})\right|_{\mathbf{r}=\mathbf{r}_i}\nonumber\\
	&=-\frac{1}{8\pi(1-\nu)(3-4\nu)\mu}\left(\frac{a}{r_{ij}}\right)^2\Big[(7-10\nu)\hat{\mathbf{r}}_{ij}\hat{\mathbf{r}}_{ij}\nonumber\\
	&\quad+2(1-2\nu)^2(\hat{\underline{\mathbf{I}}}+\hat{\mathbf{r}}_{ij}\hat{\mathbf{r}}_{ij})\Big]\cdot\mathbf{F}_i+\mathcal{O}\big(r_{ij}^{-4}\big)
\end{align}
of inclusion $i$. Via the Faxén law Eq.~(\ref{eq:Faxen_2}), we obtain its rotation
\begin{align}\label{eq:Omegai(2)_force}
	\mathbf{\Omega}_i^{(2)}&=\left.\frac{1}{2}\nabla\times\mathbf{u}_j^{(1)}(\mathbf{r})\right|_{\mathbf{r}=\mathbf{r}_i}\nonumber\\
	&=\frac{(1-2\nu)a^2}{2\pi\mu (3-4\nu) r_{ij}^3} (\hat{\mathbf{r}}_{ij}\times\mathbf{F}_i)+\mathcal{O}\big(r_{ij}^{-5}\big).
\end{align}

Finally, we sum up all contributions listed in Eqs.~(\ref{eq:Ui(0)_force}), (\ref{eq:Ui(1)_force}), and (\ref{eq:Ui(2)_force}) to find for the translation of the $i$-th inclusion
\begin{widetext}
\begin{align}\label{eq:U_total}
	\mathbf{U}_i &= \mathbf{U}_i^{(0)}+\mathbf{U}_i^{(1)}+\mathbf{U}_i^{(2)}\nonumber\\
	&= \left\{\frac{1-2(3-4\nu)\ln a}{16\pi(1-\nu)\mu}\underline{\hat{\mathbf{I}}}-\frac{1}{8\pi(1-\nu)(3-4\nu)\mu}\left(\frac{a}{r_{ij}}\right)^2\left[2(1-2\nu)^2(\hat{\underline{\mathbf{I}}}+\hat{\mathbf{r}}_{ij}\hat{\mathbf{r}}_{ij})+(7-10\nu)\hat{\mathbf{r}}_{ij}\hat{\mathbf{r}}_{ij}\right]\right\}\cdot\mathbf{F}_i \nonumber\\
	&\quad+\frac{1}{8\pi(1-\nu)\mu}\left\{\left[-(3-4\nu)\ln r_{ij} +\left(\frac{a}{r_{ij}}\right)^2\right]\hat{\underline{\mathbf{I}}}+\left[1-2\left(\frac{a}{r_{ij}}\right)^2\right]\hat{\mathbf{r}}_{ij}\hat{\mathbf{r}}_{ij}\right\}\cdot \mathbf{F}_j+\mathcal{O}\big(r_{ij}^{-4}\big).
\end{align}
Likewise, summing the contributions in Eqs.~(\ref{eq:Omegai(1)_force}) and (\ref{eq:Omegai(2)_force}), we obtain the rotation of the $i$-th inclusion
\begin{equation}\label{eq:Omega_total}
		\mathbf{\Omega}_i =\mathbf{\Omega}_i^{(1)}+\mathbf{\Omega}_i^{(2)}=-\frac{1}{4\pi\mu r_{ij}} (\hat{\mathbf{r}}_{ij}\times\mathbf{F}_j)+\frac{(1-2\nu)}{2\pi\mu (3-4\nu) }\frac{a^2}{r_{ij}^3} (\hat{\mathbf{r}}_{ij}\times\mathbf{F}_i)+\mathcal{O}\big(r_{ij}^{-5}\big).
\end{equation}

So far, we only have concentrated on two inclusions, but we can consider more inclusions using the same expressions. Each additional inclusion has the same influence on inclusion $i$ as inclusion $j$ has.
From Eqs.~(\ref{eq:matrix}) and (\ref{eq:U_total}), we identify the components of the displaceability matrices $\underline{\mathbf{M}}^{\mathrm{tt}}_{ij}$ as
\begin{equation}\label{eq:Mtt_ii}
	\underline{\mathbf{M}}^{\mathrm{tt}}_{i=j}= M_0^\mathrm{t}\left\{\big[1-2(3-4\nu)\ln a\big]\underline{\hat{\mathbf{I}}}-\sum_{\substack{k=1\\k\neq i}}^N\frac{2}{(3-4\nu)}\left(\frac{a}{r_{ik}}\right)^2\left[2(1-2\nu)^2(\hat{\underline{\mathbf{I}}}+\hat{\mathbf{r}}_{ik}\hat{\mathbf{r}}_{ik})+(7-10\nu)\hat{\mathbf{r}}_{ik}\hat{\mathbf{r}}_{ik}\right]\right\}
\end{equation}
and
\begin{equation}\label{eq:Mtt_ij}
	\underline{\mathbf{M}}^{\mathrm{tt}}_{i\neq j}=2 M_0^\mathrm{t}\left\{\left[-(3-4\nu)\ln r_{ij} +\left(\frac{a}{r_{ij}}\right)^2\right]\hat{\underline{\mathbf{I}}}+\left[1-2\left(\frac{a}{r_{ij}}\right)^2\right]\hat{\mathbf{r}}_{ij}\hat{\mathbf{r}}_{ij}\right\}+\underline{\mathbf{M}}^{\mathrm{tt}(3)}_{i\neq j},
\end{equation}
where $i,j={1,\dots,N}$ and
\begin{equation}\label{eq:M0t}
	M_0^\mathrm{t}=\frac{1}{16\pi(1-\nu)\mu}.
\end{equation}
The $\underline{\mathbf{M}}^{\mathrm{tt}(3)}_{i\neq j}$-term results from three-inclusion interactions, which we discuss in Sec.~\ref{sec:interaction_three}.
\end{widetext}

Analogously, from Eqs.~(\ref{eq:matrix}) and (\ref{eq:Omega_total}), the components of $\underline{\mathbf{M}}^{\mathrm{rt}}_{ij}$ follow as
\begin{equation}
	\underline{\mathbf{M}}^{\mathrm{rt}}_{i=j}=M_0^\mathrm{r}\frac{2(1-2\nu)a^2}{(3-4\nu) }\sum_{\substack{k=1\\k\neq i}}^N \frac{\hat{\mathbf{r}}_{ik}}{r_{ik}^3}\times\hat{\underline{\mathbf{I}}}
\end{equation}
and
\begin{equation}\label{eq:Mrt_ij}
	\underline{\mathbf{M}}^{\mathrm{rt}}_{i\neq j}= \underline{\mathbf{M}}^{\mathrm{rt}(3)}_{i\neq j} -M_0^\mathrm{r}\frac{\hat{\mathbf{r}}_{ij}}{r_{ij}}\times\hat{\underline{\mathbf{I}}}
\end{equation}
for $i,j=1,\dots,N$, together with
\begin{equation}\label{eq:M0r}
	M_0^\mathrm{r} = \frac{1}{4\pi\mu}.
\end{equation}

\subsection{Torques externally imposed or induced between the inclusions}
Now we turn from forces $\mathbf{F}_i$ and $\mathbf{F}_j$ to torques $\mathbf{T}_i$ and $\mathbf{T}_j$ acting on inclusions $i$ and $j$, respectively. For both inclusions we know the resulting rotations to lowest order, that is, in the absence of mutual interactions, from Eq.~(\ref{eq:T_Omega}). Accordingly, they read
\begin{equation}\label{eq:Omegai(0)_torque}
	\mathbf{\Omega}_i^{(0)}=\frac{1}{4\pi\mu a^2}\mathbf{T}_i
\end{equation}
and
\begin{equation}\label{eq:Omegaj(0)_torque}
	\mathbf{\Omega}_j^{(0)}=\frac{1}{4\pi\mu a^2}\mathbf{T}_j.
\end{equation}
From Eqs.~(\ref{eq:u_rot}) and (\ref{eq:T_Omega}), we also identify the undisturbed displacement fields
\begin{equation}\label{eq:ui(0)_torque}
	\mathbf{u}_i^{(0)}(\mathbf{r})=\left(\frac{a}{|\mathbf{r}-\mathbf{r}_i|}\right)^2\mathbf{\Omega}_i^{(0)}\times\left(\mathbf{r}-\mathbf{r}_i\right)
\end{equation}
and
\begin{equation}\label{eq:uj(0)_torque}
	\mathbf{u}_j^{(0)}(\mathbf{r})=\left(\frac{a}{|\mathbf{r}-\mathbf{r}_j|}\right)^2\mathbf{\Omega}_j^{(0)}\times\left(\mathbf{r}-\mathbf{r}_j\right).
\end{equation}
Similarly to Sec.~\ref{sec:interaction_two_force}, we now calculate the translation and rotation resulting directly from these displacement fields via the Faxén laws in Eqs.~(\ref{eq:Faxen_1}) and (\ref{eq:Faxen_2}), respectively, where $\mathbf{u}_j^{(0)}(\mathbf{r})$ is inserted.
We find
\begin{align}\label{eq:Ui(1)_torque}
	\mathbf{U}_i^{(1)} = \left.\left(1+\frac{a^2}{4}\nabla^2\right)\mathbf{u}_j^{(0)}(\mathbf{r})\right|_{\mathbf{r}=\mathbf{r}_i}= -\frac{1}{4\pi\mu r_{ij}}\hat{\mathbf{r}}_{ij}\times \mathbf{T}_j
\end{align}
and
\begin{align}\label{eq:Omegai(1)_torque}
	\mathbf{\Omega}_i^{(1)} =  \frac{1}{2} \left.\left(\nabla\times \mathbf{u}_j^{(0)}(\mathbf{r})\right)\right|_{\mathbf{r}=\mathbf{r}_i}=-\frac{1}{8\pi\mu r_{ij}^2}\mathbf{T}_j,
\end{align}
where we used $\mathbf{r}_{ij}\cdot \mathbf{T}_j =0$ in our geometry.
In analogy to Eq.~(\ref{eq:Sj(1)_force}), the displacement fields lead to stresslets. For inclusion $j$, it reads
\begin{align}
	\underline{\mathbf{S}}_j^{(1)}&=-\frac{2 \pi(1-\nu)\mu a^2}{(3-4\nu)}\! \left(1+\frac{a^2}{8}\nabla^2\right)\!\!\left(\frac{1}{(1-2\nu)} \hat{\underline{\mathbf{I}}} \nabla\cdot \mathbf{u}_i^{(0)}(\mathbf{r})\right.\nonumber\\	
	&\quad\left.\left.+2\left\{\nabla \mathbf{u}_i^{(0)}(\mathbf{r})+\left[\nabla \mathbf{u}_i^{(0)}(\mathbf{r})\right]^{T}\right\} \right)\right|_{\mathbf{r}=\mathbf{r}_j}\nonumber\\
	&= -\frac{2(1-\nu)}{(3-4\nu)}\frac{a^2}{r_{ij}^2}\big\{(\hat{\mathbf{r}}_{ij}\times\mathbf{T}_i)\hat{\mathbf{r}}_{ij}+\hat{\mathbf{r}}_{ij}(\hat{\mathbf{r}}_{ij}\times\mathbf{T}_i)\big\}\nonumber\\
	&\quad+\mathcal{O}\big(r_{ij}^{-4}\big).
\end{align}
$\underline{\mathbf{S}}_i^{(1)}$ follows by switching indices $i$ and $j$ in this expression. From this stresslet, we again find additional contributions to the displacement field, see Eqs.~(\ref{eq:ui(1)}) and (\ref{eq:uj(1)}). For inclusion $i$, the additional translation resulting from the stresslet-induced displacement field of inclusion $j$ is evaluated via Eq.~(\ref{eq:Faxen_1}) as
\begin{align}\label{eq:Ui(2)_torque}
	\mathbf{U}_i^{(2)} &= \left.\left(1+\frac{a^2}{4}\nabla^2\right)\mathbf{u}_j^{(1)}(\mathbf{r})\right|_{\mathbf{r}=\mathbf{r}_i}\nonumber\\
	&=-\frac{(1-2\nu)}{2\pi(3-4\nu)\mu}\frac{a^2}{r_{ij}^3}\hat{\mathbf{r}}_{ij}\times\mathbf{T}_i+\mathcal{O}\big(r_{ij}^{-5}\big).
\end{align}
Conversely, Eq.~(\ref{eq:Faxen_2}) implies
\begin{equation}\label{eq:Omegai(2)_torque}
	\mathbf{\Omega}_i^{(2)} = \mathcal{O}\big(r_{ij}^{-4}\big).
\end{equation}
Summing up the different contributions of translation in Eqs.~(\ref{eq:Ui(1)_torque}) and (\ref{eq:Ui(2)_torque}), we obtain
\begin{align}
	\mathbf{U}_i &= \mathbf{U}_i^{(1)}+\mathbf{U}_i^{(2)} \nonumber\\
	 &= -\frac{1}{4\pi\mu r_{ij}}\hat{\mathbf{r}}_{ij}\times \mathbf{T}_j\nonumber\\ &\quad-\frac{(1-2\nu)}{2\pi(3-4\nu)\mu}\frac{a^2}{r_{ij}^3}\hat{\mathbf{r}}_{ij}\times\mathbf{T}_i+\mathcal{O}\big(r_{ij}^{-5}\big).
\end{align}
From here, we read off the  entries of the $\underline{\mathbf{M}}^{\mathrm{tr}}_{ij}$ matrices as
\begin{equation}
	\underline{\mathbf{M}}^{\mathrm{tr}}_{i=j}=-\frac{2(1-2\nu)M_0^\mathrm{r}}{(3-4\nu)}\sum_{k=1,k\neq i}^N \frac{a^2}{r_{ik}^3}\hat{\mathbf{r}}_{ik}\times\hat{\underline{\mathbf{I}}}
\end{equation}
and
\begin{equation}\label{eq:Mtr_ij}
	\underline{\mathbf{M}}^{\mathrm{tr}}_{i\neq j}=\underline{\mathbf{M}}^{\mathrm{tr}(3)}_{i\neq j} - M_0^\mathrm{r}\frac{\hat{\mathbf{r}}_{ij}}{r_{ij}}\times\hat{\underline{\mathbf{I}}},
\end{equation}
where $M_0^\mathrm{r}$ is defined in Eq.~(\ref{eq:M0r}). In this expression, $ \underline{\mathbf{M}}^{\mathrm{tr}(3)}_{i\neq j}$ marks an additional three-body interaction, see Sec.~\ref{sec:interaction_three} below.

Following the analogous procedure for the rotations, we find, when summing up the contributions in Eqs.~(\ref{eq:Omegai(0)_torque}) and (\ref{eq:Omegai(1)_torque}),
\begin{align}
	\mathbf{\Omega}_i =&\ \mathbf{\Omega}_i^{(0)} + \mathbf{\Omega}_i^{(1)}\nonumber \\
	 =& \frac{1}{4\pi\mu a^2}\mathbf{T}_i-\frac{1}{8\pi\mu r_{ij}^2} \mathbf{T}_j+\mathcal{O}\big(r_{ij}^{-4}\big).
\end{align}
From here, we read off
\begin{equation}\label{eq:Mrr_ii}
	\underline{\mathbf{M}}^{\mathrm{rr}}_{i=j}= M_0^\mathrm{r}\frac{1}{a^2}\underline{\hat{\mathbf{I}}}
\end{equation}
and
\begin{equation}\label{eq:Mrr_ij}
	\underline{\mathbf{M}}^{\mathrm{rr}}_{i\neq j}= \underline{\mathbf{M}}^{\mathrm{rr}(3)}_{i\neq j}-M_0^\mathrm{r}\frac{1}{2 r_{ij}^2}\underline{\hat{\mathbf{I}}}.
\end{equation}

\section{Three-body interactions}\label{sec:interaction_three}
Using the same strategy as before, we now derive the additional expressions for three-body interactions. We keep the setting considered above for inclusions $i$ and $j$ and add an inclusion $k$ at position $\mathbf{r}_k$, subject to a force $\mathbf{F}_k$ and/or a torque $\mathbf{T}_k$. To lowest order, each inclusion produces a displacement field analogous to Eqs.~(\ref{eq:ui(0)}), (\ref{eq:uj(0)}), and (\ref{eq:Omegai(0)_torque})--(\ref{eq:uj(0)_torque}) with corresponding changes in indices.

First, we calculate the translation of inclusion $i$ due to the displacement fields that are generated by forces acting on the inclusions. The field that we use in the Faxén law Eq.~(\ref{eq:Faxen_1}) for inclusion $i$ is now a superposition of $\mathbf{u}_j^{(0)}(\mathbf{r})$ and $\mathbf{u}_k^{(0)}(\mathbf{r})$. In this way, we obtain
\begin{equation}\label{eq:Ui(1)_three}
	\mathbf{U}_i^{(1)} = \left.\left(1+\frac{a^2}{4}\nabla^2\right)\left[\mathbf{u}_j^{(0)}(\mathbf{r})+\mathbf{u}_k^{(0)}(\mathbf{r})\right]\right|_{\mathbf{r}=\mathbf{r}_i}.
\end{equation}
Analogously we calculate from Eq.~(\ref{eq:Faxen_S}) the resulting stresslet using the same superposition of displacement fields as
\begin{widetext}
\begin{align}\label{eq:Si(1)_three}
	\underline{\mathbf{S}}_i^{(1)}=&-\frac{2 \pi(1-\nu)\mu a^2}{(3-4\nu)} \left(1+\frac{a^2}{8}\nabla^2\right)\left.\left[\frac{1}{(1-2\nu)} \hat{\underline{\mathbf{I}}} \nabla\cdot \left[\mathbf{u}_j^{(0)}(\mathbf{r})+\mathbf{u}_k^{(0)}(\mathbf{r})\right]\right.\right. \nonumber \\
		&\left.\left.+2\left(\nabla \left[\mathbf{u}_j^{(0)}(\mathbf{r})+\mathbf{u}_k^{(0)}(\mathbf{r})\right]+\left\{\nabla \left[\mathbf{u}_j^{(0)}(\mathbf{r})+\mathbf{u}_k^{(0)}(\mathbf{r})\right]\right\}^{T}\right)\right]\right|_{\mathbf{r}=\mathbf{r}_i}.
\end{align}
This via Eq.~(\ref{eq:u_exp}) produces the displacement field
\begin{equation}\label{eq:ui(1)_three}
	\mathbf{u}_i^{(1)}(\mathbf{r})=-(\underline{\mathbf{S}}_i^{(1)}\cdot\nabla)\cdot\underline{\mathbf{G}}\left(\mathbf{r}-\mathbf{r}_i\right).
\end{equation}
Analogously, we obtain the corresponding expressions for inclusion $j$ and $k$ by exchanging $i$ with $j$ and $i$ with $k$, respectively.

Additional three-body contributions now arise. We focus, for instance, on the translation of inclusion $i$
\begin{equation}\label{eq:Ui(2)_three}
	\mathbf{U}_i^{(2)} = \left.\left(1+\frac{a^2}{4}\nabla^2\right)\left[\mathbf{u}_j^{(1)}(\mathbf{r})+\mathbf{u}_k^{(1)}(\mathbf{r})\right]\right|_{\mathbf{r}=\mathbf{r}_i},
\end{equation}
obtained via Eq.~(\ref{eq:Faxen_1}) from the stresslet-induced displacement fields in analogy to Eq.~(\ref{eq:ui(1)_three}). 
We concentrate, for example, in Eq.~(\ref{eq:Ui(2)_three}) on the contribution through $\mathbf{u}_k^{(1)}(\mathbf{r})$, that via the analogon of Eq.~(\ref{eq:ui(1)_three}) depends on $\underline{\mathbf{S}}_k^{(1)}$. This stresslet arises because inclusion $k$ is exposed to the displacement fields $\mathbf{u}_i^{(0)}(\mathbf{r})$ and $\mathbf{u}_j^{(0)}(\mathbf{r})$, in analogy to Eq.~(\ref{eq:Si(1)_three}). 
The field $\mathbf{u}_j^{(0)}(\mathbf{r})$ is generated by the force $\mathbf{F}_j$ acting on inclusion $j$. Thus, in reverse order, $\mathbf{F}_j$ acting on inclusion $j$ generates the displacement field $\mathbf{u}_j^{(0)}(\mathbf{r})$. Inclusion $k$ is exposed to this field and due to its rigidity generates a counterstress. In this way, the displacement field $\mathbf{u}_j^{(0)}(\mathbf{r})$ is ``reflected" by inclusion $k$ in the form of $\mathbf{u}_k^{(1)}(\mathbf{r})$. Through this chain of effects $(i\leftarrow k\leftarrow j)$,  a translation of inclusion $i$ arises, given by
\begin{align}\label{eq:Uikj(2)_force}
		\mathbf{U}_{ikj}^{(2)}
		=&-\left.\left(1+\frac{a^2}{4}\nabla^2\right)(\underline{\mathbf{S}}_k^{(1)}\cdot\nabla)\cdot\underline{\mathbf{G}}\left(\mathbf{r}-\mathbf{r}_k\right)\right|_{\mathbf{r}=\mathbf{r}_i} \nonumber\\
		=&\frac{1}{16\pi(1-\nu)(3-4\nu)\mu}\frac{a^2}{r_{ik}r_{jk}}\Big(-4(1-2\nu)\Big\{(1-2\nu)\left[(\hat{\mathbf{r}}_{ik}\cdot\hat{\mathbf{r}}_{jk})\hat{\underline{\mathbf{I}}}+\hat{\mathbf{r}}_{jk}\hat{\mathbf{r}}_{ik}\right]\nonumber\\
		&\quad+2(\hat{\mathbf{r}}_{ik}\cdot\hat{\mathbf{r}}_{jk})[\hat{\mathbf{r}}_{ik}\hat{\mathbf{r}}_{ik}+\hat{\mathbf{r}}_{jk}\hat{\mathbf{r}}_{jk}]-\hat{\mathbf{r}}_{ik}\hat{\mathbf{r}}_{jk}\Big\}+2\left[3-2\nu-4(\hat{\mathbf{r}}_{ik}\cdot\hat{\mathbf{r}}_{jk})^2\right]\hat{\mathbf{r}}_{ik}\hat{\mathbf{r}}_{jk}\Big)\cdot\mathbf{F}_j+\mathcal{O}\big((r_{ij},r_{ik})^{-4}\big).
\end{align}
This corresponds to a genuine three-body interaction.

Next, we focus on the rotations induced by these forces. The rotation $	\mathbf{\Omega}^{(1)}_{i}$ is just a superposition of the rotations induced by $\mathbf{u}_j^{(0)}(\mathbf{r})$ and $\mathbf{u}_k^{(0)}(\mathbf{r})$ via Eq.~(\ref{eq:Faxen_2}). Similarly, the next-higher order $\mathbf{\Omega}^{(2)}_{i}$ is calculated via the analogon of Eq.~(\ref{eq:Ui(2)_three}) as
\begin{equation}\label{eq:Omegai(2)_three}
	\mathbf{\Omega}_{i}^{(2)}=\left.\frac{1}{2}\nabla\times\left[\mathbf{u}_j^{(1)}(\mathbf{r})+\mathbf{u}_k^{(1)}(\mathbf{r})\right]\right|_{\mathbf{r}=\mathbf{r}_i},
\end{equation}
where the displacement fields depend on the corresponding stresslet each, see Eq.~(\ref{eq:ui(1)_three}). For example, the stresslet $\underline{\mathbf{S}}_k^{(1)}$ partly arises, because inclusion $k$ is exposed to $\mathbf{u}_j^{(0)}(\mathbf{r})$. This displacement field gets ``reflected" by inclusion $k$ due to its rigidity. This resulting $\mathbf{u}_k^{(1)}(\mathbf{r})$ via Eq.~(\ref{eq:Omegai(2)_three}) rotates inclusion $i$. This leaves us with a three-body contribution ($i\leftarrow k \leftarrow j$) to the rotation of inclusion $i$
\begin{align}\label{eq:Omegaikj(2)_force}
	\mathbf{\Omega}^{(2)}_{ikj} =& \frac{1}{2\pi (3-4\nu)\mu}\frac{a^2}{r_{jk}r_{ik}^2}\big\{(1-2\nu)\big[(\hat{\mathbf{r}}_{ik}\cdot\hat{\mathbf{r}}_{jk})(\hat{\mathbf{r}}_{ik}\times \mathbf{F}_j)+(\hat{\mathbf{r}}_{ik}\times\hat{\mathbf{r}}_{jk})(\hat{\mathbf{r}}_{ik}\cdot \mathbf{F}_j)\big]\nonumber \\
	&\quad+ 2(\hat{\mathbf{r}}_{ik}\cdot \hat{\mathbf{r}}_{jk})(\hat{\mathbf{r}}_{jk}\cdot \mathbf{F}_j)(\hat{\mathbf{r}}_{ik}\times\hat{\mathbf{r}}_{jk})\big\}+\mathcal{O}\big((r_{ij},r_{ik})^{-5}\big).
\end{align}

From Eqs.~(\ref{eq:Uikj(2)_force}) and (\ref{eq:Omegaikj(2)_force}), we can read off the additional three-body contributions $\underline{\mathbf{M}}^{\mathrm{tt}(3)}_{i\neq j}$  and $\underline{\mathbf{M}}^{\mathrm{rt}(3)}_{i\neq j}$  to Eqs.~(\ref{eq:Mtt_ij}) and (\ref{eq:Mrt_ij}),
\begin{align}
	\underline{\mathbf{M}}^{\mathrm{tt}(3)}_{i\neq j} =& M_0^\mathrm{t}\frac{1}{(3-4\nu)}\sum_{\substack{k=1\\ k\neq i,j}}^N\frac{a^2}{r_{ik}r_{jk}}\Big(-4(1-2\nu)\Big\{(1-2\nu)\left[(\hat{\mathbf{r}}_{ik}\cdot\hat{\mathbf{r}}_{jk})\hat{\underline{\mathbf{I}}}+\hat{\mathbf{r}}_{jk}\hat{\mathbf{r}}_{ik}\right]\nonumber\\
		&+2(\hat{\mathbf{r}}_{ik}\cdot\hat{\mathbf{r}}_{jk})[\hat{\mathbf{r}}_{ik}\hat{\mathbf{r}}_{ik}+\hat{\mathbf{r}}_{jk}\hat{\mathbf{r}}_{jk}]-\hat{\mathbf{r}}_{ik}\hat{\mathbf{r}}_{jk}\Big\} +2\left[3-2\nu-4(\hat{\mathbf{r}}_{ik}\cdot\hat{\mathbf{r}}_{jk})^2\right]\hat{\mathbf{r}}_{ik}\hat{\mathbf{r}}_{jk}\Big)
\end{align}
and
\begin{equation}
\underline{\mathbf{M}}^{\mathrm{rt}(3)}_{i\neq j} = \frac{2 M_0^\mathrm{r}}{(3-4\nu)}\sum_{\substack{k=1\\ k\neq i,j}}^N\frac{a^2}{r_{jk}r_{ik}^2}\left\{(1-2\nu)\left[(\hat{\mathbf{r}}_{ik}\cdot\hat{\mathbf{r}}_{jk})(\hat{\mathbf{r}}_{ik}\times \hat{\underline{\mathbf{I}}})+(\hat{\mathbf{r}}_{ik}\times\hat{\mathbf{r}}_{jk})(\hat{\mathbf{r}}_{ik}\cdot \hat{\underline{\mathbf{I}}})\right]+ 2(\hat{\mathbf{r}}_{ik}\cdot \hat{\mathbf{r}}_{jk})(\hat{\mathbf{r}}_{ik}\times\hat{\mathbf{r}}_{jk})(\hat{\mathbf{r}}_{jk}\cdot \hat{\underline{\mathbf{I}}})\right\}.
\end{equation}
Here we used again the definitions of $M_0^\mathrm{t}$ in Eq.~(\ref{eq:M0t}) and $M_0^\mathrm{r}$ in Eq.~(\ref{eq:M0r}).

Turning to torques $\mathbf{T}_i$, $\mathbf{T}_j$, and $\mathbf{T}_k$ acting on inclusions $i$, $j$, and $k$, respectively, instead of forces, three-body interactions arise as well. The corresponding contributions to the translation and rotation of inclusion $i$ are again obtained from Eqs.~(\ref{eq:Ui(2)_three}) and (\ref{eq:Omegai(2)_three}). Part of the displacement field $\mathbf{u}_k^{(1)}(\mathbf{r})$ results in analogy to Eq.~(\ref{eq:ui(1)_three}), because the rigidity of inclusion $k$ leads in analogy to Eq.~(\ref{eq:Si(1)_three}) to a stresslet $\underline{\mathbf{S}}_k^{(1)}$. This stresslet partly arises, because inclusion $k$ is exposed to the displacement field $\mathbf{u}_j^{(0)}(\mathbf{r})$ now generated by the torque $\mathbf{T}_j$ acting on inclusion $j$, see Eqs.~(\ref{eq:Omegaj(0)_torque}) and (\ref{eq:uj(0)_torque}). Up to our desired order, we find three-body contributions ($i \leftarrow k \leftarrow j$) to the translation due to the torque $\mathbf{T}_j$ of the form
\begin{align}\label{eq:Uikj(2)_torque}
		\mathbf{U}_{ikj}^{(2)}
		=&-\frac{1}{2\pi(3-4\nu)\mu}\frac{1}{r_{ik}r_{jk}^2}\left\{(1-2\nu)\left[(\hat{\mathbf{r}}_{ik}\cdot\hat{\mathbf{r}}_{jk})\hat{\underline{\mathbf{I}}}\times\hat{\mathbf{r}}_{jk}+\hat{\mathbf{r}}_{jk}\hat{\mathbf{r}}_{ik}\times\hat{\mathbf{r}}_{jk}\right]\right.\\
		&\left.+2(\hat{\mathbf{r}}_{ik}\cdot\hat{\mathbf{r}}_{jk})\hat{\mathbf{r}}_{ik}\hat{\mathbf{r}}_{ik}\times\hat{\mathbf{r}}_{jk}\right\}\cdot\mathbf{T}_j+\mathcal{O}\big((r_{ij},r_{ik})^{-5}\big),
\end{align}
but not to the rotation, because
\begin{equation}\label{eq:Omegaikj(2)_torque}
	\mathbf{\Omega}_{ikj}^{(2)} = \mathcal{O}\big((r_{ij},r_{ik})^{-4}\big).
\end{equation}
Therefore, we can formulate from Eqs.~(\ref{eq:Uikj(2)_torque}) and (\ref{eq:Omegaikj(2)_torque}) the remaining three-body contributions in Eqs.~(\ref{eq:Mtr_ij}) and (\ref{eq:Mrr_ij}) as
\begin{align}
\underline{\mathbf{M}}^{\mathrm{tr}(3)}_{i\neq j} &= -\frac{2 M_0^\mathrm{r}}{(3-4\nu)}\sum_{\substack{k=1 \\ k\neq i,j}}^N\frac{1}{r_{ik}r_{jk}^2}\left\{(1-2\nu)\left[(\hat{\mathbf{r}}_{ik}\cdot\hat{\mathbf{r}}_{jk})\hat{\underline{\mathbf{I}}}\times\hat{\mathbf{r}}_{jk}+\hat{\mathbf{r}}_{jk}\hat{\mathbf{r}}_{ik}\times\hat{\mathbf{r}}_{jk}\right]+2(\hat{\mathbf{r}}_{ik}\cdot\hat{\mathbf{r}}_{jk})\hat{\mathbf{r}}_{ik}\hat{\mathbf{r}}_{ik}\times\hat{\mathbf{r}}_{jk}\right\}
\end{align}
and
\begin{equation}
\underline{\mathbf{M}}^{\mathrm{rr}(3)}_{i\neq j} = \underline{\mathbf{0}}.
\end{equation}
\end{widetext}

\section{Removing the logarithmic divergence for vanishing net force}\label {sec:ln-problem}
We now return to the logarithmic divergence mentioned in Secs.~\ref{sec:intro} and \ref{sec:Green} that arises in two spatial dimensions. Specifically, we note the contribution $\sim\ln r$ in the Green's function in Eq.~(\ref{eq:Greens}). Generally, this logarithmic divergence carries over to the displacement fields generated by net forces acting on the inclusions, see, for example, Eq.~(\ref{eq:u_trans}). 

For $N$ inclusions, we write the overall displacement field as
\begin{equation}
	\mathbf{u}(\mathbf{r})=\sum_{i=1}^N	\mathbf{u}_i(\mathbf{r}).
\end{equation}
Considering for our present purpose only  the logarithmic terms, we are according to Eqs.~(\ref{eq:Greens}) and (\ref{eq:u_trans}) left with
\begin{align}\label{eq:u_ln}
	\mathbf{u}(\mathbf{r})&=\frac{-(3-4\nu)}{8\pi(1-\nu)\mu}\sum_{i=1}^N	\ln(|\mathbf{r}-\mathbf{r}_i|)\mathbf{F}_i+\dots \nonumber\\
	&=\frac{-(3-4\nu)}{8\pi(1-\nu)\mu}\bigg[\ln(|\mathbf{r}-\mathbf{r}_j|)\mathbf{F}_j\nonumber\\
	&\quad+\sum_{\substack{i=1 \\ i\neq j}}^N	\ln(|\mathbf{r}-\mathbf{r}_i|)\mathbf{F}_i\bigg]+\dots \ ,	
\end{align}
where $j\in\{1, \dots , N\}$.
Vanishing net force on the whole collection of discrete inclusions implies for the $j$-th inclusion
\begin{equation}\label{eq:no_net_force}
	\mathbf{F}_j=-\sum_{\substack{i=1 \\ i\neq j}}^N\mathbf{F}_i.
\end{equation}
Inserting Eq.~(\ref{eq:no_net_force}) into Eq.~(\ref{eq:u_ln}), we obtain
\begin{align}\label{eq:end}
	\mathbf{u}(\mathbf{r})&=\frac{-(3-4\nu)}{8\pi(1-\nu)\mu}\sum_{\substack{i=1 \\ i\neq j}}^N	\mathbf{F}_i\ln\left(\frac{|\mathbf{r}-\mathbf{r}_i|}{|\mathbf{r}-\mathbf{r}_j|}\right)+\dots\ .	
\end{align}
This displacement field correctly tends to $\mathbf{0}$ for $|\mathbf{r}|\rightarrow\infty$.

Overall, we find that, if we assume vanishing net force on the whole set of inclusions, we actually do not observe the divergence of the displacement field at large distances. Thus at least two inclusions are necessary to remove the divergence problem. For just one inclusion exposed to a net force, the divergence remains.

Physically, this implies that, if the inclusions do not interact with the outside world but only amongst each other and elastically trough the membrane, Newton's third law guarantees that the displacement field remains finite. The same argument applies for any collection of force centers acting on the elastic membrane. It for $\nu\rightarrow1/2$ likewise describes the behavior of two-dimensional incompressible fluid films under low-Reynolds-number conditions. Since, according to Eq.~(\ref{eq:end}), only ratios of distances enter the logarithm in the end, there is no inconsistency with apparently dimensionful quantities as arguments of the logarithm in the Green's function as might have been suspected from the notation in Eq.~(\ref{eq:Greens}). 

In contrast, if net forces do act from outside onto the inclusions in two-dimensional systems, long-ranged interactions with the lateral boundaries of the membrane or thin film emerge. These interactions with the boundaries arise, no matter how far away the boundaries are from the inclusions. Such boundaries are present in any realistic setup. In a corresponding mathematical description, the boundary conditions then need to be included into the formalism, using a different Green's function associated with this different setup.

\section{Comparison between two- and three-dimensional setups}\label{sec:2Dvs3D}
In Refs.~\onlinecite{puljiz2017pre} and \onlinecite{puljiz2019pre-higher-order} spherical inclusions embedded in three-dimensional elastic media were described. That situation represents the three-dimensional counterpart to our two-dimensional setup. It is obviously of interest to compare the results for the different dimensionalities.  For general statements, we consider the leading orders of the $r$-dependencies of different quantities.

First, a few quantities do not depend on the distance. These are the translation $\mathbf{U}^{(0)}_i$ induced by $\mathbf{F}_i$ in Eq.~(\ref{eq:Ui(0)_force}) and the rotation $\mathbf{\Omega}^{(0)}_i$ induced by $\mathbf{T}_i$ in Eq.~(\ref{eq:Omegai(0)_torque}). Differences only arise for the prefactors. These relations are reflected by the corresponding matrix entries $\underline{\mathbf{M}}^{\mathrm{tt}}_{i=j}$ and $\underline{\mathbf{M}}^{\mathrm{rr}}_{i=j}$ in Eqs.~(\ref{eq:Mtt_ii}) and (\ref{eq:Mrr_ii}), respectively. Thus, for these relations, there does not exist any difference in the $r$-dependency between two- and three-dimensional setups.

The situation changes for the Green's function $\underline{\mathbf{G}}(\mathbf{r})$ in Eq.~(\ref{eq:Greens}). In the two-dimensional case it shows a leading logarithmic dependency, while it features a leading inverse $r$-dependency $\sim 1/r$ in the three-dimensional case \cite{puljiz2017pre,puljiz2019pre-higher-order}. The same is found for the leading orders of the displacement fields $\mathbf{u}^{(0)}_i(\mathbf{r})$ induced by $\mathbf{F}_i$, see Eq.~(\ref{eq:ui(0)}). Combining this conclusion with our analysis in Sec.~\ref{sec:ln-problem}, we note a central difference between two- and three-dimensional setups. In two dimensions, individual inclusions that are subject to a net force do interact with each other through the surrounding medium, no matter how far apart they are from each other. Only if the net force on a group of inclusions vanishes, we can neglect their influence on the surrounding medium with increasing distance from the group. The situation is manifestly different in three dimensions. There, the influence of individual inclusions subject to net forces decays with increasing distance from the inclusions.

Concerning further differences, we note that the stresslet $\underline{\mathbf{S}}^{(1)}$ in Eq.~(\ref{eq:Sj(1)_force}) shows a dependency $\sim 1/r$, while the corresponding three-dimensional expression features a dependency $\sim 1/r^2$ \cite{puljiz2017pre,puljiz2019pre-higher-order}. The same applies to the displacement field $\mathbf{u}^{(0)}_i(\mathbf{r})$ induced by $\mathbf{T}_i$, see Eq.~(\ref{eq:ui(0)_torque}) as against Refs.~\onlinecite{puljiz2017pre,puljiz2019pre-higher-order}. Naturally, these differences affect the higher orders as well when they are tracked through the presented formalism.

\begin{figure}
	\includegraphics[width=\linewidth]{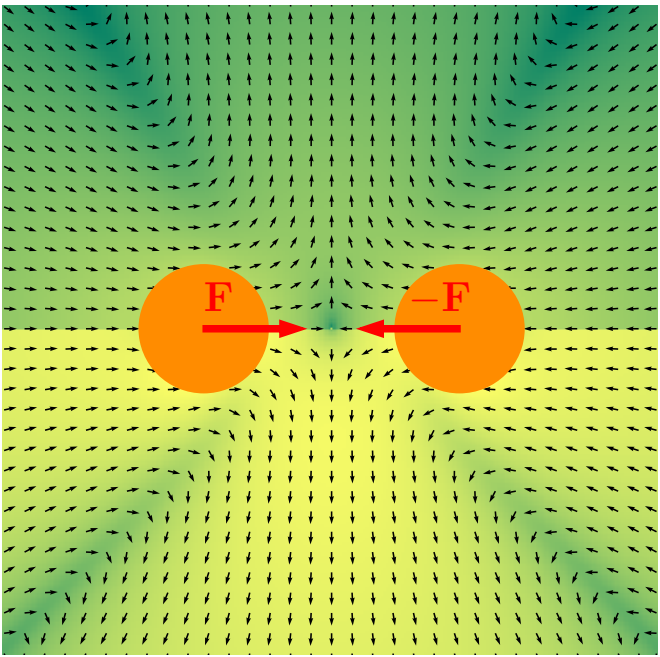}
	\caption{Illustration of the in-plane displacement field that two inclusions mutually attracting each other by the indicated forces $\pm\mathbf{F}$ generate. In the top half of the figure, the three-dimensional case is depicted using the corresponding formulas listed in Ref.~\onlinecite{puljiz2017pre}. In three dimensions, the inclusions represent rigid spheres and we confine ourselves to a plane that contains both centers of these spheres. In the bottom half, we show the results for our two-dimensional case, where the inclusions represent rigid disks. For better visibility, local directions of the displacement field are indicated by small dark arrows of identical length. The local magnitudes of the displacement field are represented by the color code on a logarithmic scale. Brighter colors mark larger amplitudes of displacement. We set the Poisson ratio to $\nu=1/2$.}
\end{figure}

\begin{figure}
	\includegraphics[width=\linewidth]{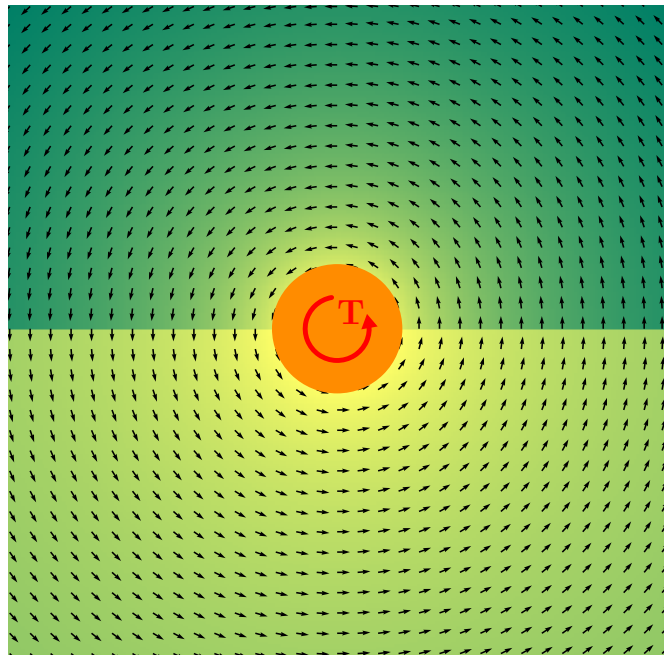}
	\caption{Illustration of the displacement field in a three- and two-dimensional setting generated by one rigid inclusion exposed to a net torque $\mathbf{T}$. The representation is analogous to Fig.~1.}
\end{figure}

For further comparison, we illustrate for two basic setups the differences in the displacement fields concerning the two- and three-dimensional case. First, we consider a situation of two inclusions mutually attracting each other by forces $\mathbf{F}$ and $-\mathbf{F}$, see Fig.~1. The net force vanishes, so that there is no logarithmic divergence in the displacement field. We compare the resulting two-dimensional case in the bottom half of Fig.~1 to the three-dimensional case in the top half for an incompressible system ($\nu=1/2$). The magnitude of the displacement field is color coded on a logarithmic scale, and the brighter color in the bottom half indicates larger amplitudes of displacement in the two-dimensional case. In analogy to that, Fig.~2 shows the corresponding comparison for a torque $\mathbf{T}$ applied to a rigid inclusion. The local direction of the displacement field (see the normalized small arrows) is not influenced by the dimensionality. Yet, the brighter color in the bottom half again indicates that the magnitude of the displacement field is larger in the two-dimensional situation.

\section{Conclusions}\label{sec:conclusion}

In the present work, we demonstrated that also in the two-dimensional case the theoretical characterization of interactions between rigid inclusions embedded in an elastic or fluid environment is well defined for infinitely extended systems. This statement applies as long as the overall net force acting on the inclusions vanishes. Accordingly, we derived the explicit analytical expressions for translational and rotational couplings between the inclusions as mediated by the elastic or fluid environment. 
While compressible embedding media are addressed in a linearly elastic case, low-Reynolds-number flows of incompressible liquids are covered for fluid surroundings. 

As already explained, the two-dimensional treatment includes the three-dimensional bulk situation for systems that are homogeneous and largely extended along the third dimension. Thus, our rigid disks in this case actually represent infinitely extended, aligned cylinders. For example, we may consider two long, parallel, conducting wires of circular cross section, pierced through the bulk of a soft elastic gel. If an electric direct current runs through these wires, they either attract or repel each other, depending on whether the current runs into the same or opposite direction in the two wires, respectively \cite{jackson-book}. Assuming a very soft gel, its elastic modulus can for instance be as low as $1~$Pa \cite{huang2016buckling}. We further assume the gel to be incompressible. Moreover, we approximate the conducting cores of the wires by infinitely thin conducting lines when calculating their mutual electromagnetic interaction. If the wires are both subject to an electric direct current of $20~$A, if they together with their insulation feature a radius of $1.5~$mm, and if their center-to-center distance is $1~$cm, our evaluations indicate a change in distance between the wires of $1.75~$mm. This effect thus becomes visible, and even the displacements of the elastic gel could be visualized by embedding a few smaller tracer particles. As also mentioned before, the characterization of thin linearly elastic membranes can be reduced to a two-dimensional framework as well \cite{landau1986theory}. 

Particularly, the description directly applies to thin elastic or fluid membranes or films that by themselves are isotropic in the in-plane directions, are approximately incompressible along the normal direction, and contain inclusions that are roughly disk-like within the range of the membrane. Several example systems feature the approximate incompressibility along the normal. Importantly, this applies to lipid bilayers that form the basis of the outer membrane of many types of biological cells and cell organelles or vesicles \cite{noguchi2001self, edidin2003lipids, reynwar2007aggregation, sakuma2010pore}, although inclusions can bend the membrane. In that case, curvature needs to be taken into account.

Concerning the more macroscopic scale, we mention free-standing thin films or bubbles of smectic A liquid crystals as fluid systems \cite{eremin2011two, may2014freely}. The smectic layers typically extend along the in-plane film directions, while on average the liquid crystalline molecules are oriented along the normal. Thus, the fluid along the in-plane directions appears isotropic. Conversely, compression along the layer normal is usually hindered in smectic liquid crystals \cite{degennes1993physics}, as the molecular layers would need to be driven into or separated from each other. This supports our requirement of approximate incompressibility along the normal direction. An analogous situation for elastic realizations emerges for smectic liquid crystal elastomers. There, likewise, the elastic modulus for compression and dilation along the layer normal is significantly increased \cite{nishikawa1997smectic}. 

It will be inspiring to analyze several of these example systems in the future when they are functionalized by inclusions, now that we have the appropriate formalism at hand. Moreover, actuation and activation processes may be facilitated in this way. To this end, we envisage an extension to thin sheets and membranes composed of viscoelastic materials \cite{puljiz2019pre-visco, richter2021epl}. Another extension concerns the dynamic coupling of deformations of thin elastic membranes to flows in surrounding fluids \cite{PhysRevE.100.032610,daddi-moussa-ider_lisicki_gekle_2017}.

\begin{acknowledgments}
The authors thank the German Research Foundation (Deutsche Forschungsgemeinschaft) DFG for support through the research grant no.\ ME 3571/5-1. Moreover, A.M.M.\ acknowledges support by the DFG through the Heisenberg Grant no.\ ME 3571/4-1. 
\end{acknowledgments}

%


\end{document}